\documentclass[floatfix,twocolumn,aps,superscriptaddress,nofootinbib,pra,10pt]{revtex4-2}

\usepackage[american]{babel}
\usepackage{amsmath,amsthm,amssymb,amsfonts, mathtools}
\usepackage{stmaryrd}
\usepackage[colorlinks=true,linktocpage=true]{hyperref}
\hypersetup{
    allcolors  = {blue},
}

\usepackage{xcolor}
\usepackage[all]{xy}

\usepackage{graphicx}

\usepackage{braket}
\usepackage{float}
\usepackage{array}

\allowdisplaybreaks

\def\emph{\textbf}

\theoremstyle{plain}
\newtheorem{theorem}             {Theorem}

\newtheorem{lemma}      [theorem]{Lemma}

\newtheorem*{theorem*}    {Theorem}
\newtheorem*{proposition*}{Proposition}
\newtheorem*{lemma*}      {Lemma}
\newtheorem*{corollary*}  {Corollary}
\newtheorem*{conjecture*} {Conjecture}

\theoremstyle{definition}

\newtheorem*{definition*}{Definition}
\newtheorem*{example*}   {Example}
\newtheorem*{question*}   {Question}
\newtheorem*{idea*}  {Idea}

\theoremstyle{remark}

\usepackage{proof}

\usepackage{colonequals}

\def\implies{\Rightarrow}

\usepackage{graphicx}
\usepackage{multirow}
\usepackage{url}
\usepackage{tcolorbox}

\usepackage{bbold}

\usepackage[normalem]{ulem}

\usepackage{tikz}
\usetikzlibrary{quantikz}

\pgfdeclarelayer{edgelayer}
\pgfdeclarelayer{nodelayer}
\pgfsetlayers{edgelayer,nodelayer,main}

\usepackage{tikz}

\pgfdeclarelayer{edgelayer}
\pgfdeclarelayer{nodelayer}
\pgfsetlayers{edgelayer,nodelayer,main}

\tikzstyle{none}=[inner sep=0pt]
\tikzstyle{copy}=[fill=white, draw=black, shape=circle]
\tikzstyle{new}=[-, thick, draw={rgb,255: red,81; green,41; blue,241}]
\tikzstyle{new_dashed}=[-, thick, dashed, draw={rgb,255: red,81; green,41; blue,241}]
\tikzstyle{alice}=[-, fill={rgb,255: red,74; green,201; blue,255}, draw={rgb,255: red,12; green,60; blue,216}]
\tikzstyle{bob}=[-, fill={rgb,255: red,255; green,123; blue,125}, draw={rgb,255: red,171; green,18; blue,21}]
\tikzstyle{alice_dashed}=[-, dashed, draw={rgb,255: red,74; green,201; blue,255}]
\tikzstyle{bob_dashed}=[-, dashed, draw={rgb,255: red,255; green,123; blue,125}]
\usepackage{tikz}
\tikzstyle{copy}=[fill=white, draw=black, shape=circle]
\tikzstyle{gn}=[rectangle, rounded corners=0.8em, fill={rgb,255: red,216; green,248; blue,216}, draw=black, line width=0.8 pt, inner sep=3pt, minimum width=1.5em, minimum height=1.5em]
\tikzstyle{rn}=[rectangle, rounded corners=0.8em, fill={rgb,255: red,232; green,165; blue,165}, draw=black, line width=0.8 pt, inner sep=3pt, minimum width=1.5em, minimum height=1.5em]
\tikzstyle{hadamard}=[fill=yellow, draw=black, shape=rectangle]
\pgfdeclarelayer{edgelayer}
\pgfdeclarelayer{nodelayer}
\pgfsetlayers{edgelayer,nodelayer,main}

\usepackage{soul}
\newcommand{\inlshort}{INL -- International Iberian Nanotechnology Laboratory, Braga, Portugal}

\newcommand{\uffshort}{Instituto de F\'{i}sica, Universidade Federal Fluminense, Niter\'{o}i -- RJ, Brazil}

\newcommand{\cfumshort}{Centro de F\'{i}sica, Universidade do Minho, Braga, Portugal}

\newcommand{\cfupshort}{Departamento de F\'{i}sica e Astronomia, Faculdade de Ciências, Universidade do Porto, Porto, Portugal.}

\begin{document}
\title{Non-stabilizerness and entanglement from cat-state injection}

\author{Filipa C. R. Peres}
\email{filipa.peres@inl.int}
\affiliation{\inlshort}%\affiliation{\inl}
\affiliation{\cfupshort}%\affiliation{\cfum}

\author{Rafael Wagner}
\email{rafael.wagner@inl.int}
\affiliation{\inlshort}%\affiliation{\inl}
\affiliation{\cfumshort}%\affiliation{\cfum}

\author{Ernesto F. Galv\~ao}
%\email{ernesto.galvao@inl.int}
\affiliation{\inlshort}%\affiliation{\inl}
\affiliation{\uffshort}%\affiliation{\uff} 

\date{\today}

\begin{abstract}
Recently, cat states have been used to heuristically improve the runtime of a classical simulator of quantum circuits based on the diagrammatic ZX-calculus. Here we investigate the use of cat-state injection within the quantum circuit model. We explore a family of cat states, $\left| \mathrm{cat}_m^* \right>$, and describe circuit gadgets using them to concurrently inject non-stabilizerness (also known as magic) and entanglement into any quantum circuit. We provide numerical evidence that cat-state injection does not lead to speed-up in classical simulation. On the other hand,  we show that our gadgets can be used to widen the scope of compelling applications of cat states. Specifically, we show how to leverage them to achieve savings in the number of injected qubits, and also to induce scrambling dynamics in otherwise non-entangling Clifford circuits in a controlled manner.
\end{abstract}

\maketitle

\section{Introduction}\label{sec: intro}

It is often the case that, in order to precisely capture the physics behind quantum information tasks or quantum computation protocols, one must treat the same phenomena through various lenses. To consider a concrete example, Shor's algorithm might seem simple, but it is only once we understand the physics behind it in terms of computational complexity~\cite{Shor1999}, nonclassical resources~\cite{Ahnefeld2022coherence}, and hardware implementation~\cite{MartinLopez2012,Vandersypen2001}, among others, that we can properly glimpse the enormous challenges behind implementing this algorithm. This plurality of techniques and their benefits push different communities to be in constant contact and discussion. 

A powerful outcome of such a collaborative effort between quantum information scientists, category theory mathematicians, and computer scientists was the development of ZX-calculus, a formal diagrammatic calculus for quantum theory ~\cite{coecke2017picturing,wetering2020zxcalculus,heunen2019categories}. This diagrammatic treatment is now being applied in various research fields such as  quantum error correction~\cite{deBeaudrap2020zxcalculusis,bombin2023unifying} and classical simulation of quantum computation~\cite{Kissinger2020reducing, Kissinger+2022,wetering2022classically}. Moreover, research into the latter field has also seen extensive progress in the last years~\cite{BG2016, BSS2016, BBCCGH2019, Qassim2021, Kissinger+2022}. As a striking example of the importance of bridging the research of different communities, in Ref.~\cite{Kissinger+2022}, Kissinger \textit{et. al} showed how the so-called cat states, introduced as a tool in~\cite{Qassim2021}, can be extremely powerful devices for improving the runtime of classical simulation of quantum circuits within the diagrammatic ZX-calculus.

In our work, motivated by this recent application of cat states, we choose these states as our case study to analyze improvements within three different topics of research. To that end, first, we define a family of cat states that we characterize in terms of non-stabilizerness (commonly referred to as magic) and entanglement. These cat states are closely related to a family of magic states investigated in \cite{BeveCHK2019} for their relationship to catalysis and their provably optimal synthesis over the $\left< CX,\,T \right>$ gate set. Next, we describe a family of quantum circuit gadgets that fault-tolerantly inject non-stabilizer and entangling unitaries into quantum circuits, via these resource states. This family of gadgets allows us to highlight interesting applications of cat states. Specifically, we analyze (i) the power of cat-state injection for classical simulation of quantum computation within the quantum circuit model framework; (ii) the possible reductions in the quantum resources needed for certain quantum computations; and (iii) how the injected unitaries introduce complexity to the task of classically simulating quantum dynamics. For the latter application, we use a simple test-bed for hardness of simulation that is the evidence of scrambling of quantum information in hard-to-simulate quantum circuits.

The structure of our work is as follows. In Sec.~\ref{sec: Background}, we introduce the basics of gadgets in quantum circuits, present an overview of classical simulation, revisit various aspects of the quantum resources we will study in cat states, and review the essentials of doped quantum circuits and scrambling of information. In Sec.~\ref{sec: Results}, we present our results. We explore and characterize a family of cat states, construct gadgets that inject magic and entangling unitaries into any quantum circuit provided we have access to such states, and discuss the three aforementioned applications of these gadgets. Finally, we provide some concluding remarks in Sec.~\ref{sec: Discussion}.

\section{Background}\label{sec: Background}

\subsection{Fault-tolerant quantum computation and circuit gadgets}\label{subsec: Background - gadgets}

The thirst for fault-tolerant quantum computing has driven a lot of research into error-correcting codes wherein redundancy is used to reduce or eliminate the nefarious effect of noise. More specifically, these error-correcting codes work by encoding logical qubits into several physical qubits. After that, the computation is driven by operations, such as gates and measurements, applied to the encoded qubits. The simplest way of acting with a gate $G$ on an encoded qubit $q_L$ would be to simply act with that same gate independently on each of the physical qubits of $q_L$. In a given code, gates that admit such an implementation are said to be \textit{transversal}~\cite{NielsenChuang}. Unfortunately, it has been shown that no useful error-correcting code admits a universal set of transversal gates~\cite{EastinKnill2009}.

Stabilizer codes are a particular type of error-correcting code wherein the encoding and decoding can be accomplished using circuits with Clifford gates only. Moreover, whilst Clifford gates admit transversal implementations, any non-Clifford gate, required to unlock universality~\cite{DiVincenzo1995}, will be non-transversal. For this reason, Clifford gates are commonly regarded as \textit{free} operations -- in the sense that they can be implemented easily/cheaply -- whereas any non-Clifford gate is hard/expensive to implement. 

The magic-state injection model of quantum computation, proposed by Bravyi and Kitaev in 2005~\cite{BravyiKitaev2005}, addresses the aforementioned difficulties. In their seminal work, the authors describe a 15-to-1 magic-state distillation protocol wherein a single, low-noise copy of the non-stabilizer resource state $\left| T \right> \coloneqq \left( \left| 0\right> + e^{i\pi /4}\left| 1\right> \right)/\sqrt{2}$ can be extracted from $15$ noisier copies of the same state, via the application of only (cheap) Clifford gates. The distilled resource state can then be used to implement the non-Clifford $T$ gate, $T \coloneqq \mathrm{diag}\left( 1, e^{i\pi /4}\right),$ using only Clifford unitaries and classical feedforward. Thus, in this model, quantum computations are carried out by \textit{adaptive} Clifford circuits with magic-state qubits obtained via a magic-state distillation protocol.

The adaptive stabilizer circuit constructions used to (fault-tolerantly) implement non-stabilizer gates via the injection of magic resource states are known as \textit{gadgets}. Besides the paradigmatic example of the $T$ gate discussed above, there are several other well-known gadgets implementing gates such as the Toffoli gate, the controlled-$S$ gate, and gates in higher levels of the Clifford hierarchy~\cite{zhou2000methodoly}. In complete generality, the existence of gadgets for elements in the third level of the Clifford hierarchy is guaranteed by the following theorem.

\begin{theorem}[Adapted from~\cite{zhou2000methodoly}]
Let $U$ be an $n$-qubit gate in the third level of the Clifford hierarchy. Then, the state $\vert \psi \rangle = U\vert 0 \rangle^{\otimes n}$ can be prepared fault-tolerantly by using only (possibly adaptive)  gates and measurements in the second level of the Clifford hierarchy.
\label{Theorem: Fault tolerance 3rd level}
\end{theorem}

In our work, we will analyze a family of magic and entangled states that we obtain from the \textit{original cat states} defined in~\cite{Qassim2021}, and which we call here \textit{star cat states}. In particular, we describe the corresponding non-stabilizer and entangling unitaries that can be injected into Clifford circuits if we have access to a cat state factory and explicitly describe the gadgets that can be used to perform the injection. We call these \textit{cat gadgets}. As we will show later, Theorem~\ref{Theorem: Fault tolerance 3rd level} theoretically guarantees that cat gadgets exist and can be implemented in a fault-tolerant manner. 

\subsection{Classical simulation of quantum computation}\label{subsec: Background - Classical simulation}
In parallel to the research into fault tolerance, the field of classical simulation of quantum circuits has also seen significant research over the past two decades. This is because improving classical simulators allows us to effectively probe the threshold between classical and quantum computation, i.e., to establish which quantum circuits effectively bring forth quantum computational advantage. Moreover, classical simulation aids in the development of hardware and software (e.g. error-correcting codes) and gives insights into new, quantum-inspired classical algorithms.

Since non-stabilizer operations upgrade Clifford circuits into universal quantum computation, the \textit{amount} of magic in a quantum circuit -- usually quantified via \textit{magic measures} or \textit{magic monotones} (cf. Sec.~\ref{subsec: Background: Resource Theory of Magic}) -- provides a way of quantifying the hardness of classical simulation. A thorough description of the connection between different magic monotones and the cost of classical simulation is provided in~\cite{Seddon+2021}. Here, we refrain from going into such extensive technical details and instead focus only on the results of immediate relevance to our work.

Recent work in stabilizer-rank simulators has given rise to classical simulation algorithms that scale polynomially with the number $n$ of qubits in the quantum circuit, but exponentially with the number $t$ of non-Clifford gates. As explained above, a unitary circuit with $t$ non-Clifford gates can be transformed into an adaptive Clifford circuit with a given input magic state, $\left| \Psi_{R} \right>$. For instance, in the case where all non-Clifford gates are $T \coloneqq \text{diag}(1,e^{i\pi/4})$ gates, the input resource state is simply the tensor product of $t$ copies of the single-qubit magic state $\left| T \right> \coloneqq \left( \left| 0\right> + e^{i\pi /4}\left| 1\right> \right)/\sqrt{2}$, i.e., $\left| \Psi_{R} \right> \coloneqq \left| T \right>^{\otimes t}$. Such a state can be decomposed as:
\begin{equation}
    \left| T \right>^{\otimes t} = \sum_{i=1}^r c_i \left| \phi_i \right>,
    \label{eq: stabilizer decomposition}
\end{equation}
where $c_i$ are complex coefficients and the sum extends over $r$ $t$-qubit stabilizer states $\left| \phi_i \right>.$

Having decomposed the input  resource state  into a linear combination of stabilizer states, it is possible to efficiently compute the output state of each of the $r$ adaptive Clifford circuits~\cite{BG2016}. Weighting each of these states by the proper coefficient allows us to reconstruct the output state of the computation and compute the expectation value $\left< O \right>$ of any observable $O$ of interest.

The runtime of these stabilizer-rank simulators scales with $\mathcal{O}\left(r^2\right)$~\cite{BG2016} such that their inefficiency is actually hidden in $r$, which scales exponentially with the number of non-Clifford gates in the quantum circuit. Therefore, finding decompositions of the form of Eq.~\eqref{eq: stabilizer decomposition} with minimal $r$ has a direct impact on the simulation cost. The stabilizer rank, $\chi\left( \left| \Psi_{R} \right>\right)$, of a given non-stabilizer state $\left| \Psi_{R} \right>$, is a magic monotone defined as the minimum value of $r$. Therefore, the best runtime of stabilizer-rank simulators is given by $\mathcal{O}\left( \chi\left( \left| \Psi_{R} \right>\right)^2 \right),$ for a quantum circuit with input state $\left| \Psi_{R} \right>$.

Observing that a single copy of the state $\left| T \right>$ has stabilizer rank $\chi\left( \left| T \right>\right) = 2,$ and because the stabilizer rank is submultiplicative, i.e., $\chi\left( \left| \Psi_{R}\right> \otimes \left| \Phi_{R} \right>\right) \leq \chi\left( \left| \Psi_{R} \right>\right)\chi\left( \left| \Phi_{R} \right>\right)$, one arrives at the conclusion that the runtime of these algorithms is (at most) $\mathcal{O}\left( 2^t \right).$ Fortunately, it is possible to do better. In particular, Qassim \textit{et al.}~\cite{Qassim2021} show that $\chi\left( \left| T \right>^{\otimes 6}\right) = 6,$ so that using such decomposition yields a scaling $\mathcal{O}\left( 6^{t/6} \right)=\mathcal{O}\left( 2^{\alpha t} \right),$ with $\alpha = \log_2(6)/6 \approx 0.4308.$

In fact, the best  currently known value for this coefficient is $\alpha = 0.3963$ obtained asymptotically in Ref.~\cite{Qassim2021} and via a partial decomposition provided in Ref.~\cite{Kissinger+2022}. 

\subsubsection{Cat states and improved classical simulation costs}

To obtain the best upper bound for the cost of classical simulation known to date when using a stabilizer-rank approach, Ref.~\cite{Qassim2021} resorted to a family of magic states called \textit{cat states} and defined as
\begin{equation}
    \left| \mathrm{cat}_m \right> = \frac{1}{\sqrt{2}} \left( \left| T \right>^{\otimes m} + \left| T ^{\perp}\right>^{\otimes m}\right)\,,
    \label{eq: definition of cat states}
\end{equation}
where $\left| T ^{\perp}\right>$ denotes the state orthogonal to $\left| T \right>$.  Besides being non-stabilizer states, it is clear that these cat states are also entangled. 

Qassim \textit{et al.}~\cite{Qassim2021} noted that these states are low-rank magic states such that:
\begin{equation}
    \frac{\chi\left( \left| T \right>^{\otimes m}\right)}{2} \leq \chi\left( \left| \mathrm{cat}_m \right>\right) \leq \chi\left( \left| T \right>^{\otimes m}\right).
    \label{eq: bounds on rank of cats}
\end{equation}
Using this result, they searched for decompositions of $\left| \mathrm{cat}_m \right>$ (rather than $\left| T \right>^{\otimes m}$) with larger $m$ than is feasible for tensor products of $\left| T \right>.$ Finding the decompositions of these cat states then allowed them to set better upper bounds for $\chi\left(\left| T \right>^{\otimes m}\right).$ 

Building upon these observations, Ref.~\cite{Kissinger+2022} proposed a heuristic approach for improving the runtime of classical simulation using ZX-calculus. (For a brief description of the ZX-calculus framework see Appendix~\ref{app: ZX-calculus basics}. We recommend Ref.~\cite{wetering2020zxcalculus} for a more in-depth discussion.) Essentially, their proposal works as follows:
\begin{enumerate}
    \item[(a)] Take the quantum circuit that we want to simulate;
    \item[(b)] Transform it, using ZX simplification rules, into a graph-like ZX-diagram (see Appendix~\ref{app: ZX-calculus basics} or Ref.~\cite{wetering2020zxcalculus});
    \item[(c)] Simplify the diagram as much as possible; 
    \item[(d)] Search the diagram for cat-state-like sub-diagrams and use the low-rank stabilizer decompositions of such states to break the large diagram into a (small) number of other diagrams;
     \item[(e)] Return to step (c) and carry out this procedure iteratively until all diagrams have been reduced to constants that can be used to compute the desired expectation value.
\end{enumerate}
By searching first for structures that can be associated with cat states, the procedure ensures that the multiplicity of diagrams increases more slowly than it would if we were to consider the best decomposition of $\left| T \right>^{\otimes m}$ states. Because this procedure is heuristic, the promised asymptotic scaling is still the same, i.e. $\mathcal{O}\left( 2^{0.3963 t}\right)$, however, in practice, the whole simulation is much faster, and quantum circuits that were previously out of reach for other classical simulators are simulable using this technique~\cite{Kissinger+2022,wetering2022classically}. 

Here, we point out that this procedure enables fast classical simulation at the cost of moving away from the circuit model, which is a natural and familiar representation of quantum computations. Because of that, we lose some insight into which features of the original circuit are allowing the simulation speed-up.

\subsection{Resource theory of magic}\label{subsec: Background: Resource Theory of Magic}

As explained in the previous section, magic is often used to quantify the cost of classical simulation of quantum dynamics, introducing an exponential overhead. We can treat magic formally as a resource using the resource theory of magic. In a quantum resource theory~\cite{ChitambarGour2019}, one describes sets of free states and free quantum channels and attempts to quantify the \textit{amount} of a given resource using the so-called monotones, i.e., quantities that can only monotonically decrease upon application of free operations.

The resource theory of magic considers states that can be prepared by Clifford circuits as free objects, and any Clifford operation as a free channel. In such a way, one can define various monotones, i.e., functions of the form $M: \mathcal{F} \to [0,1]$ such that $M(C(\rho)) \leq M(\rho)$ for $C$ some Clifford unitary. A first example is that of the stabilizer rank $\chi$, defined in the previous subsection. Another one is the robustness of magic (RoM), denoted as $\mathcal{R}$. Letting $\mathcal{S}_n := \{\vert \phi_i \rangle\}_i$ be the set of $n$-qubit pure stabilizer states, we have that~\cite{HowardCampbel2017}
\begin{equation}
    \mathcal{R}(\rho) := \min_x \{\sum_i \vert x_i \vert \, : \, \rho = \sum_i x_i \vert \phi_i \rangle \langle \phi_i \vert \}.
    \label{eq: RoM optimization problem}
\end{equation}
In Ref.~\cite{HowardCampbel2017}, the authors make a comprehensive study of how this monotone can be used in many different ways, including (i) characterizing the cost of classical simulation using a quasi-probability approach, (ii) bounding the number of $T$ gates needed to synthesize a certain (non-stabilizer) unitary, etc. They also compute the RoM of multiple copies of the $\left| T \right>$ state, as well as of other magic states such as $\left| CS \right> \coloneqq CS \left| + \right>^{\otimes 2} = \left( 1,1,1,i\right)/2$, $\left| CCZ \right>\coloneqq CCZ \left| + \right>^{\otimes 3} = \left( 1,1,1,1,1,1,1,-1\right)/\sqrt{8}$, and $\left| \mathrm{Hoggar} \right>\coloneqq \left( 1+i, 0, -1, 1, -i, 1, 0, 0\right)/\sqrt{6}$. Their results are summarized in Table~\ref{tab: RoM from Howard and Campbell}. We use $\left|  \Phi_{max} \right>$ to denote the maximally robust state of 2 qubits which has $\mathcal{R}\left( \left| \Phi_{max} \right> \right) = \sqrt{5}$ (cf.~\cite{HowardCampbel2017} for details on the state), while $\left| CS \right>$ is the maximally robust 2-qubit state with all equal weights. 

The Hoggar state is the 3-qubit state with maximum robustness~\cite{HowardCampbel2017}. Independently of the ensuing discussion and work, it is worth noting that while this observation makes this state special in a sense, to the best of our knowledge, a simple gadget that fault-tolerantly injects the corresponding unitary $U_{\mathrm{Hog.}}$ (generating this state from the state $\left| 0 \right>^{\otimes 3}$) into a quantum circuit is unknown. This may stem from the fact that $U_{\mathrm{Hog.}}$ is not in the third level of the Clifford hierarchy, so that Theorem \ref{Theorem: Fault tolerance 3rd level} does not apply.

\begin{table}[t]
    \centering
    \begin{tabular}{>{\centering\arraybackslash}p{0.7cm} p{1.2cm} >{\centering\arraybackslash}p{4.1cm}}
        \hline\hline
        $n$ & $\left| \Psi \right>$ & $\mathcal{R}\left( \left| \Psi \right> \right)$~\cite{HowardCampbel2017}\\
         \hline
         1  & $\left| T \right>$    & $\sqrt{2} \approx 1.4142$ \\
         \hline
         \multirow{3}{*}{2}  & $\left| T \right>^{\otimes 2}$  & $\left( 1+3\sqrt{2}\right)/3 \approx 1.7476$ \\
           & $\left| CS \right>$  & 2.2 \\
           & $\left| \Phi_{max} \right>$  & $\sqrt{5} \approx 2.2361$ \\
         \hline
         \multirow{3}{*}{3}  & $\left| T \right>^{\otimes 3}$  & $\left(1+4\sqrt{2}\right)/3 \approx 2.2190$ \\
           & $\left| CCZ \right>$ & 2.555 \\
           & $\left| \mathrm{Hoggar} \right>$  & 3.8 \\
           \hline
         4  & $\left| T \right>^{\otimes 4}$    & $\left( 3 + 8\sqrt{2} \right)/5 \approx 2.8627$ \\
         \hline\hline
    \end{tabular}
    \caption{\textbf{Robustness of magic of different magic states as determined in}~\cite{HowardCampbel2017}\textbf{.} The first column highlights the number of qubits of the state, the second column specifies the state under consideration, and the last column indicates the corresponding robustness of magic.}
    \label{tab: RoM from Howard and Campbell}
\end{table}

\subsection{Resource theory of entanglement}\label{subsec: Background - RT entanglement}

In contrast to the relatively recent resource theory of magic discussed above, the resource theory of entanglement has its roots in the early development of quantum information, with many important contributions over a long period ~\cite{amico2008entanglement,horodecki2009entanglement,guhne2009entanglement,eisert2010area,friis2018entanglement}. This resource theory can be analyzed from two main perspectives: a dynamical one, where we consider the entanglement properties of quantum channels and unitary gates~\cite{eisert2021entangling,gour2020dynamical,theurer2020quantifying}, or a statical one, where the entanglement present in bipartite or, more broadly, multipartite quantum states is the focus of examination~\cite{horodecki2009entanglement,ChitambarGour2019,walter2016multipartite}.

In the static formalism, the resource theory considers the free states to be all separable states, and the free operations to be either (i) resource non-generating, in which case they are termed separability-preserving or non-entangling operations~\cite{lami2023nosecondlaw}, (ii) local operations with classical communication (LOCC)~\cite{bennett1996concentrating}, or (iii) local operations with shared randomness (LOSR)~\cite{buscemi2012all}. Ref.~\cite{vedral1997quantifying} prescribed minimal requirements for entanglement monotones. The theory for such monotones is extremely rich, and still under development, especially for the case of multipartite entanglement.

We will be interested in multipartite entanglement quantification. This kind of entanglement becomes intrinsically complicated to quantify in a consistent manner due to the exponentially increasing number of nonequivalent entanglement classes under LOCC. It is common that, for studying multipartite entanglement, one often wants ways to distinguish between degrees of entanglement, such as \textit{bipartite} versus \textit{tripartite} entanglement, etc.  In the case of the present work, we are solely interested in showing that we have some degree of entanglement in a state and that this resource is injected into a circuit, generating entangling unitaries. For that, we will use the quantifier of multipartite entanglement introduced by Meyer and Wallach~\cite{meyer2002global,brennen2003anobservable} that is simple to calculate and interpret. Denoting  $\mathcal{D}(\mathcal{H})$ as the set of all quantum states acting over space $\mathcal{H}$, this monotone has the nice feature of being easy to compute for a given $m$-partite pure state $\vert \psi \rangle \in \mathcal{D}(\mathcal{H}_1 \otimes \dots \otimes \mathcal{H}_m)$:
\begin{equation}\label{eq: ent monotone}
    \mathcal{E}\left(\vert \psi \rangle \right) := 2\left(1 - \frac{1}{m}\sum_{k=0}^{m-1}\text{Tr} \left(\rho_k^2\right)\right),
\end{equation}
where $\rho_k$ describes the state of a subsystem, given by $\rho_k = \text{Tr}_{\setminus \mathcal{H}_k}(\vert \psi \rangle \langle \psi \vert)$, that is, the partial trace of $\vert \psi \rangle$ with respect to all subsystems but $\mathcal{H}_k$.

The quantity $\mathcal{E}$ is an entanglement quantifier that is (i) normalized, i.e. $\forall \vert \psi \rangle,\, \mathcal{E}(\vert \psi \rangle) \in [0,1]$, (ii) faithful, $\mathcal{E}(\vert \psi \rangle)=0 \iff \vert \psi \rangle$ is separable (that is, a product state), (iii) invariant under local unitary operations, (iv) maximal for (balanced) Greenberger-Horne-Zeilinger (GHZ) states, i.e., $\mathcal{E}(\vert \text{GHZ}_m \rangle)=1, \forall m$, (v) monotonic under LOCC.

\subsection{Doping random quantum circuits with magical resources}\label{subsec: Background - Doping}

Distinct quantum computing implementations have different limitations on the set of gates that can be natively applied. For instance, computers based on dual-rail encoding using linear-optical components easily implement any single-qubit unitary, while entangling operations are fairly costly due to the probabilistic nature of Bell measurements~\cite{Browne2005,Grice2011,Ewert2014}. Other physical platforms have non-Clifford operations as the decisive factor in increasing the simulation overhead. 

Universal random quantum circuits have properties that are provably hard to simulate classically; specifically, they are known to be chaotic and one cannot avoid an overhead that is  exponential in the $T$-count~\cite{Seddon+2021,BG2016,HowardCampbel2017,Leone2021quantumchaos} (recall Sec.~\ref{subsec: Background - Classical simulation}). The complexity gap between the efficient classical simulation of Clifford unitary circuits, and that of generic unitaries suggests it may be instructive to study the controlled introduction of  non-Clifford gates into structured circuits. To that end, a framework known as \textit{doped random circuits} proved to be useful in gradually obtaining unitary designs via magic doping~\cite{Haferkamp2022, Haferkamp2020homeopathy}, and in capturing the role of magic in quantum chaos~\cite{Oliviero2022blackhole, Oliviero2022measuring, Leone2021quantumchaos, Leone2022learning, Leone2022retrieving, ahmadi2022quantifying} and many-body entanglement spectrum statistics~\cite{Zhou2020entanglementspectrum, True2022transitions}.

\begin{figure}[h]
    \centering
    \includegraphics[width=\columnwidth]{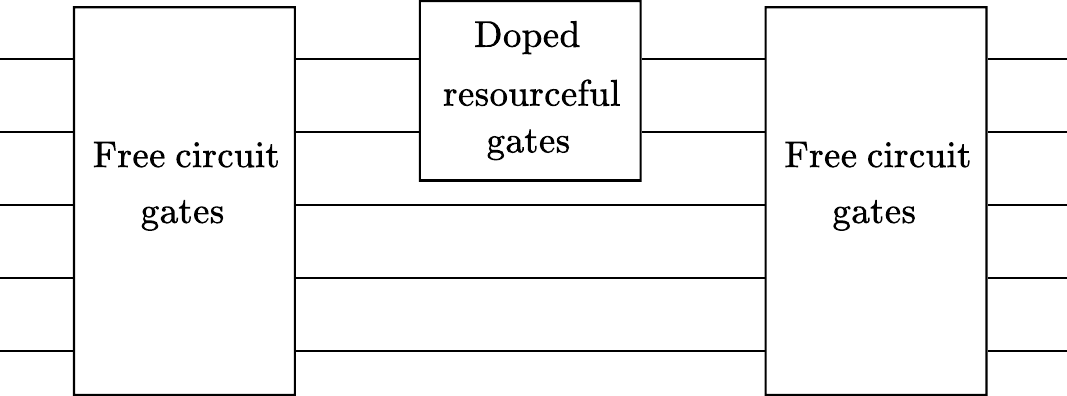}
    \caption{\textbf{Doping resources into quantum circuits.} Generic blocks of free circuits, e.g. Clifford circuits or separable qubit  operations, receive controlled doses of resourceful gates, e.g. magic gates in Clifford+$T$ architectures, entangling gates in photonics, or Hadamard gates in the Feynman-path approach to classical simulation of quantum computation.}
    \label{fig:doped}
\end{figure}

So far, the literature on doped circuits has focused mainly on one model of hardness of computation based on magic. Here we consider a more general approach to  doped quantum circuits, as captured by Fig.~\ref{fig:doped}. We consider doping a circuit with resourceful dynamics in-between blocks of free gates. The procedure of applying (randomly in a circuit) resourceful gates in-between free circuits  may involve multiple slices of the construction depicted in Fig.~\ref{fig:doped}.

This more general view of doped circuits allows us to consider paradigms other than the injection of magic since, broadly speaking, other gates might be considered resourceful in different  settings. An interesting example is classical simulation using a Feynman-path approach~\cite{AaronsonChen2017}, in which only gates that do not preserve the computational basis incur in an exponential overhead to the cost of simulation. In linear optical quantum computation, separable (qu\textit{dit}) gates are considered free, and we may dope the system with resourceful entangling gates, such as the \textsc{cnot} gate.

The most studied doped random circuits are constituted by single-qubit non-Clifford gates (e.g. $T$ gates or $T^\dagger$ gates) evolved adjointly by Clifford unitaries~\cite{Leone2021quantumchaos}. In our work, we will consider the generic case of doping non-stabilizer and entangling unitaries into non-entangling Clifford circuits. This is natural for the analysis of cat states that, besides being magical, are also multipartite entangled (cf. Sec.~\ref{subsec: Background - Classical simulation}).

We shall see that both non-stabilizerness and entanglement contribute to the simulation complexity when we inject cat states into non-entangling Clifford circuits. Moreover, it provides a natural set of resource states for studying chaotic dynamics, for which both magic and entanglement  are necessary. Resource theories based on the complexity associated with quantum information scrambling were recently introduced~\cite{Halpern2022uncomplexity, Garcia2022RTScrambling}. Here we will be interested in estimating numerically how many, and what type, of cat states need to be injected into a non-entangling Clifford circuit to achieve quantum information scrambling, a topic we review next.

\subsubsection{Scrambling and quantum chaos}

As already mentioned, one direct application of doped Clifford circuits involves understanding the role of magic in quantum chaos generated by random circuits. Quantum chaos is a surprising property that some dynamical quantum systems exhibit, and it is assumed to be deeply connected with many of the most intriguing problems in physics, such as the information loss paradox~\cite{Hayden2007,Lashkari2013}.

Scrambling of quantum information and quantum chaos are viewed as two related but different phenomena~\cite{xu2020doesscrambling}. A well-established witness of scrambling of information, known as the out-of-time-order correlator (OTOC)  $F(\tau)$ is defined as
\begin{equation}\label{eq: OTOC}
    F(\tau) := \text{Tr}\left[W(\tau)^\dagger V^\dagger W(\tau)V\rho \right]\,,
\end{equation}
where $V,W(0)$ are two local operators such that $[V,W(0)] = 0$, $\rho$ is the initial state of the system, and $W(\tau) = U^\dagger(\tau)W(0)U(\tau)$ corresponds to the dynamical evolution of $W(0)$ in the Heisenberg picture, with $U$ denoting the unitary implementing the scrambling dynamics.  This quantity can be estimated in various ways~\cite{Halpern2018quasiprobability,Wagner2023circuits}, and since it is in general complex-valued, we normally treat its real $\mathfrak{R}[F(t)]$ and imaginary parts $\mathfrak{I}[F(t)]$ separately.

Heuristically, one witnesses scrambling typical from non-integrable dynamics when~\cite{GonzalezOTOC2019,Halpern2018quasiprobability}:  
\begin{enumerate}
    \item[(a)] there is an exponential decay of the real part of the OTOC.
    \item[(b)] there is no recovery of the real part of the OTOC.
\end{enumerate}
Exponential decay (a) indicates that initially localized information is being spread to many degrees of freedom through entanglement generation. Lack of recovery (b) is a typical heuristics characterizing the lack of integrability of the dynamics~\cite{GonzalezOTOC2019,Halpern2018quasiprobability}, or also the presence of non-Clifford resources~\cite{Mi2021}. Recurrence times for non-integrable dynamics are exponentially large as compared to the integrable case~\cite{Hosur2016,Bocchieri1957quantum}. We will see this exact behavior -- expected to be hard to simulate classically~\cite{prosen2007istheefficiency} -- in our numerical studies, when investigating the effect of cat state injection into random quantum circuits for the generation of  scrambling dynamics, as captured by the OTOC in Eq.~\eqref{eq: OTOC}. This allows for a study of the concurrent roles of non-stabilizerness and entanglement in the complexity of classical simulation.

\begin{figure*}[t]
    \centering
    \includegraphics[width=1\textwidth]{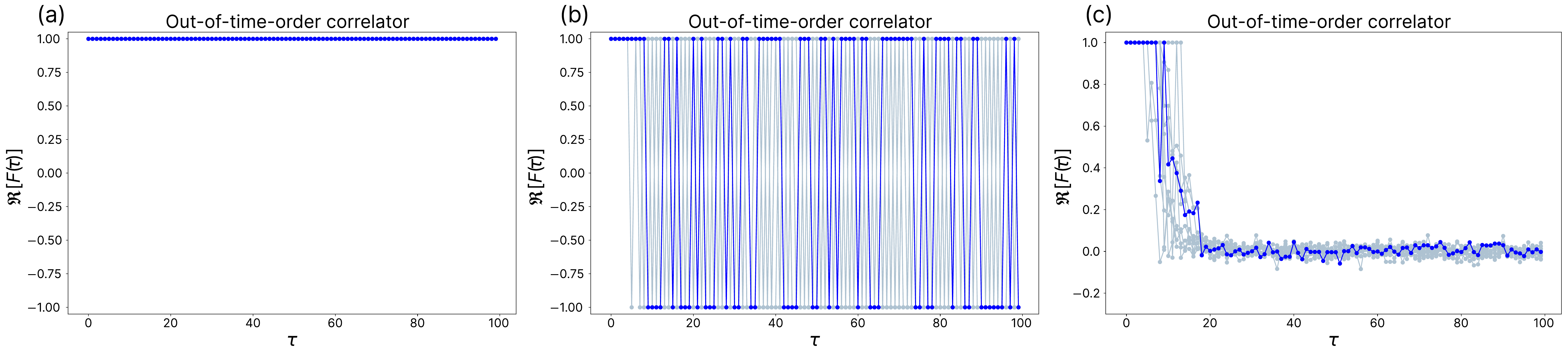}
    \caption{\textbf{Scrambling arising from ideal universal random quantum circuits.} For a quantum circuit on $n=10$ qubits, we numerically simulate the OTOC, $F(\tau)$, for $V =  \mathbb{1}^{\otimes (n-1)}\otimes X$ and $W(0) = Z \otimes \mathbb{1}^{\otimes (n-1)}$. The unitary $U_\tau$ is generated by random and dense blocks $U_{B_i}$, $0\leq i \leq 100$ of gates of $10$ layers each. We take each block to represent a time step $\tau$ and consider the total unitary to be the concatenation of all the blocks, e.g., at time $\tau = 2$ we have $U_{2} = U_{B_2}\circ U_{B_1}$ and at time $\tau=3$ we generate a new block unitary $U_{B_3}$ so that $U_3 = U_{B_3}\circ U_{B_2}\circ U_{B_1}$, and so on up to $\tau=100$. We choose gates from the sets (a) non-entangling Clifford, i.e. $\{H,S\}$, (b) Clifford $\coloneqq \{H,S,CX\}$, and finally (c) Clifford+$T \coloneqq \{H,S,CX,T\}$, all acting on an initial state  $\rho = \vert 0 \rangle \langle 0 \vert$. We present $10$ instances of this evolution and highlight the last of these. In sub-figure (c), the exponential decrease of the real part of the OTOC, together with the lack of revival, are signatures of non-integrable dynamics.}
    \label{fig: Ideal scrambling}
\end{figure*}

In Fig.~\ref{fig: Ideal scrambling} we illustrate these ideas by numerically simulating scrambling dynamics in ideal random quantum circuits. We choose a 10-qubit system evolved according to unitaries described by the gate sets: Fig.~\ref{fig: Ideal scrambling}(a) non-entangling Clifford $\coloneqq\{H,S\}$, Fig.~\ref{fig: Ideal scrambling}(b) Clifford $\coloneqq \{H,S,CX\}$ and  Fig.~\ref{fig: Ideal scrambling}(c) of Clifford+$T\coloneqq\{H,S,CX,T\}$, applied uniformly at random. Controlled-\textsc{not} gates are applied between any pair of qubits, regardless of the distance between them. We plot $10$ instances showing that, for each case, the behavior of the OTOC is generic, and highlight the last instance as a probe of that behavior. We show many instances, instead of using an averaging method, since it was recently argued~\cite{kobrin2023comment} that averaging over instances might lead to misleading conclusions regarding emergent thermalization/scrambling in many-body dynamics.

It is important to note that the kind of (ideal) dynamics presented in Fig.~\ref{fig: Ideal scrambling}(c) is believed to be both hard to learn and simulate classically, for any choice of classical simulation paradigm~\cite{prosen2007istheefficiency,Leone2021quantumchaos,Leone2022learning}. We shall see later how cat states can be injected into non-entangling, random Clifford circuits, moving the behavior of the dynamics from Fig.~\ref{fig: Ideal scrambling}(a) towards (c), by simultaneously injecting non-stabilizerness and entanglement resources, and hence introducing universality into the random circuit.

\section{Results}\label{sec: Results}

As explained in the previous section, the original cat states defined in Eq.~\eqref{eq: definition of cat states} have been exploited in two main contexts. In Ref.~\cite{Qassim2021}, they were used as a convenient tool for finding lower-rank decompositions of $\left| T \right>^{\otimes n}$ leading to the improvements on the upper bounds of the cost of classical simulation of quantum computation with magic states $\left| T \right>^{\otimes n}$. On the other hand, Ref.~\cite{Kissinger+2022} explores the decompositions of cat states directly in order to achieve a heuristic improvement of their (strong) classical simulator using ZX-calculus. 

Motivated and inspired by these works, here we aim to explore resource states that are both magic and entangled, expanding their realm of applicability beyond the classical simulation of quantum circuits. To that end, we start by defining and characterizing an infinite family of cat states, $\left| \mathrm{cat}^*_m \right>$. Then, we construct an explicit magic state injection gadget, which we dub \textit{cat gadget}, that uses these \textit{star} cat states to inject a family of $m$-qubit unitaries $V_m\,$. 

One reason for exploring this family of star cat states (rather than using the original one) is the recognition that $\left| \mathrm{cat}^*_m \right>$ has the same RoM as $\left| \mathrm{cat}_{m+1} \right>$ whilst having one less qubit. As a consequence, these two states hold the potential of injecting unitaries with the same number of $T$ gates. This will be discussed further in due time.

We exploit three main applications for our cat gadget. First, in a more direct connection with the prior works~\cite{Qassim2021} and~\cite{Kissinger+2022}, we investigate if the classical simulation of quantum circuits can be sped up by the search for patterns $V_m$, in analogy to the ZX simulator reviewed in Sec.~\ref{subsec: Background - Classical simulation}. Then, we capitalize on the properties of $\left| \mathrm{cat}^*_m \right>$ states by (i) analyzing how the distillation of these states has the potential for resource reduction in actual quantum computations; and (ii) showing how the injection of (only) these states into otherwise non-entangling Clifford circuits leads to the appearance of scrambling, which, as we discussed, is associated with quantum computational advantage.

\subsection{The family of star cat states}\label{subsec: Results - characterizing new cat states}

In Sec.~\ref{subsec: Background - Classical simulation}, we presented the definition of the original cat states as given by Eq.~\eqref{eq: definition of cat states}. Moreover, we explained that these states are such that $\chi \left( \left| T^{\otimes m} \right> \right) \leq 2 \chi \left( \left| \mathrm{cat}_m \right> \right),$ which enabled finding better upper bounds for the cost of (strong) classical simulation~\cite{Qassim2021} and led also to a heuristic runtime improvement of the classical ZX simulator~\cite{Kissinger+2022}.

Let us now introduce the family of star cat states which we will denote $\left| \mathrm{cat}_m^{*} \right>$. Assuming access to a factory of the original cat states given by Eq.~\eqref{eq: definition of cat states}, we obtain the star cat states $\left| \mathrm{cat}_m^{*} \right>$ by measuring one qubit of the original cat state in the Pauli-$Y$ basis, and post-selecting on outcome $+1$:
\begin{equation}
    \left| \mathrm{cat}_m^{*} \right>\left| +_i \right> \propto \left( I^{\otimes m} \otimes \left| +_i \right> \left< +_i \right| \right) \left| \mathrm{cat}_{m+1} \right>\,.
    \label{eq: new family of cat states}
\end{equation}
Dropping the measured qubit, we are left with the state $\left| \mathrm{cat}_m^{*} \right>$ which has one less qubit than the original cat state.

It is fairly easy to check that $\left| CS \right> \equiv \left| \mathrm{cat}_2^{*} \right>,$ which is to say that $\left| CS \right>$ can be obtained from $\left| \mathrm{cat}_3 \right>$ simply by measuring one of its qubits in the appropriate basis. This observation will be relevant in Sec.~\ref{subsec: Results - constructing the cat gadget} when we construct the cat gadgets. 

In the description above we obtained the star cat state by postselecting on a $+1$ outcome of the $Y$ measurement. It is easy to check that a $-1$ outcome also results in the desired star cat state, if we apply a $Z^{\otimes m}$ correction on the unmeasured qubits:
\begin{equation*}
    \left| \mathrm{cat}_m^{*} \right>\left| -_i \right> \propto \left( Z^{\otimes m} \otimes \left| -_i \right> \left< -_i \right| \right) \left| \mathrm{cat}_{m+1} \right>\,.
\end{equation*}

Before proceeding, a small note on terminology is in order. From now on, we use the term ``\textit{original} cat states'' to refer to the states specified in Eq.~\eqref{eq: definition of cat states}, and either ``\textit{star} cat states'' or simply ``cat states'' to refer to the family defined via Eq.~\eqref{eq: new family of cat states}.

Whilst recognizing that the family of star cat states can be obtained from the original one simply via a stabilizer measurement will be useful in the ensuing discussion, explicit expressions of $\left| \mathrm{cat}_m^{*} \right>$ can also be provided so that:
\begin{equation*}
    \left| \mathrm{cat}_m^{*} \right> = \frac{1}{\sqrt{2}} \left( \left| \mathrm{cat}_m \right> + e^{-i \pi / 4} \left| \mathrm{cat}_m^{\perp} \right>\right)\,,
\end{equation*}
where $\left| \mathrm{cat}_m^{\perp} \right> = \left( \left| T \right>^{\otimes m} - \left| T^{\perp} \right>^{\otimes m} \right)/\sqrt{2}\,.$ Alternatively, we can recognize that the equation above leads to the following form for these non-stabilizer and entangled resource states:
\begin{equation*}
    \left| \mathrm{cat}_m^{*} \right> = \frac{1}{2^{m/2}} \sum_{s\in \mathbb{F}_2^m} i^{\left\lfloor \left| s \right|/2 \right\rfloor}\left| s \right>\,,
\end{equation*}
where the sum is taken over all the $m$-bits binary bit strings, $\left| s \right|$ denotes the string's Hamming weight, and $\left\lfloor a \right\rfloor$ denotes the floor function which yields the last integer smaller or equal to $a$.

In Ref. \cite{BeveCHK2019}, $\left| W_n \right>$ states -- which are Clifford-equivalent to our $\left| \mathrm{cat}_n^* \right>$ -- are mentioned in the context of resource state conversion with and without catalysis (see Figure 3 therein). Thus, the unitaries injected by our star cat states, depicted below in Fig.~\ref{fig: V_m family construction}, closely resemble the unitaries depicted in Fig. 4 of \cite{BeveCHK2019}.

\subsubsection{Non-stabilizerness}

Since all we have done was to apply a stabilizer operation, concretely, a Pauli measurement, it is clear that $\chi \left( \left| \mathrm{cat}_m^{*} \right> \right) \leq \chi \left( \left| \mathrm{cat}_{m+1} \right> \right).$

Eq.~\eqref{eq: bounds on rank of cats} gives us an upper bound on the stabilizer rank of the original cat states: $\chi \left( \left| \mathrm{cat}_m \right> \right) \leq  \chi \left( \left| T^{\otimes m} \right> \right).$ Therefore, we immediately note that
\begin{equation}
    \chi \left( \left| \mathrm{cat}_m^{*} \right> \right) \leq \chi \left( \left| \mathrm{cat}_{m+1} \right> \right) \leq \chi \left( \left| T^{\otimes (m+1)} \right> \right).
    \label{eq: chi relationship between all states}
\end{equation}

We are also interested in characterizing the robustness of magic of this family of states. There is a result in~\cite{HowardCampbel2017} that states that if the $n$-qubit unitary used to obtain a certain resource state, $\left| \Psi_{R} \right>$, from the state $\left| + \right>^{\otimes n}$ cannot be synthesized with $t$ (or fewer) $T$ gates then
\begin{equation*}
    \mathcal{R} \left( \left| T \right>^{\otimes t} \right) < \mathcal{R}\left( \left| \Psi_{R} \right) \right> < \mathcal{R} \left( \left| T \right>^{\otimes (t+1)} \right)\,.
\end{equation*}

For the case of cat states, this translates into the following bounds for the RoM:
\begin{equation}
    \mathcal{R} \left( \left| \mathrm{cat}^{*}_m \right> \right) \leq \mathcal{R}\left( \left| \mathrm{cat}_{m+1} \right) \right> < \mathcal{R} \left( \left| T \right>^{\otimes (m+1)} \right)\,,
    \label{eq: RoM relationship between all states}
\end{equation}
which orders these states by the RoM in the same way as Eq.~\eqref{eq: chi relationship between all states} orders them by the stabilizer rank.

In Table~\ref{tab: RoM of cat states}, we present the values of the RoM of both the original and the star cat states  which we computed by solving the optimization problem set in Eq.~\eqref{eq: RoM optimization problem}. For comparison, the RoM of tensor products of the $\left| T \right>$ state is also presented (as calculated in~\cite{HowardCampbel2017}). We call attention to two facts regarding these results: first, for all the values of $m$ presented, $\mathcal{R}\left( \left| \mathrm{cat}_m^{*} \right> \right) = \mathcal{R}\left( \left| \mathrm{cat}_{m+1} \right> \right)$; second,
\begin{equation}
    \mathcal{R} \left( \left| T \right>^{\otimes m} \right) < \mathcal{R}\left( \left| \mathrm{cat}_m^{*} \right> \right) < \mathcal{R} \left( \left| T \right>^{\otimes (m+1)} \right)\,
\end{equation}
which means that the family of unitaries, $V_m:$ $V_m \left| + \right>^{\otimes m} = \left| \mathrm{cat}_m^{*} \right>$ cannot be synthesized with fewer than $(m+1)$ $T$ gates. In fact, Lemma A.7. of \cite{BeveCHK2019} informs us that these unitaries are precisely synthesized with $(m+1)$ $T$ gates, a fact which holds for any $m$. It is reasonable to believe that the first observation will also be valid for any $m$.
\begin{table}[t]
    \centering
    \begin{tabular}{>{\centering\arraybackslash}p{0.6cm} >{\centering\arraybackslash}p{2cm} >{\centering\arraybackslash}p{2cm} >{\centering\arraybackslash}p{2.6cm}}
        \hline\hline
        $m$ & $\mathcal{R}\left( \left| \mathrm{cat}_m^{*} \right> \right)$ & $\mathcal{R}\left( \left| \mathrm{cat}_{m+1} \right> \right)$ & $\left| T \right>^{\otimes (m+1)}$~\cite{HowardCampbel2017}\\
        \hline
        2 & 2.2 & 2.2 & 2.2190 \\
        3 & 2.55556 & 2.55556 & 2.8627 \\
        4 & 3.65625 & 3.65625 & 3.68705 \\
        \hline\hline
    \end{tabular}
    \caption{\textbf{Robustness of magic of the star and original cat states.} To more easily verify that Eq.~\eqref{eq: RoM relationship between all states} holds, the values of the RoM of tensor products of $\left| T \right>$, calculated in~\cite{HowardCampbel2017}, are also presented.}
    \label{tab: RoM of cat states}
\end{table}

In Sec.~\ref{subsec: Results - constructing the cat gadget}, we will discuss in a bit more detail the shape of the family of unitaries $V_m$ defined in the previous paragraph, and show how they can be injected into a given random circuit as long as we have access to a factory of $\left| \mathrm{cat}^{*}_{m} \right>$ states.

It may seem curious that the measurement of one qubit of $\left| \mathrm{cat}_m \right>$ in the Pauli-$Y$ basis preserves the RoM of the states. In the following, we analyze the consequences of performing alternative Pauli measurements, illustrating that magic may or may not be preserved under different measurements. For instance, we could measure one qubit of $\left| \mathrm{cat}_m \right>$ in the eigenbasis of $X$ which, assuming the outcome $+1$, leads to the family of states given by
\begin{equation*}
    \left| \phi_m^{*} \right>\left| + \right> \propto \left( I^{\otimes m} \otimes \left| + \right> \left< + \right| \right) \left| \mathrm{cat}_{m+1} \right>\,.
\end{equation*}
Alternatively, performing a $Z$ measurement with outcome $+1$ yields
\begin{equation*}
    \left| \varphi_m^{*} \right>\left| 0 \right> \propto \left( I^{\otimes m} \otimes \left| 0 \right> \left< 0 \right| \right) \left| \mathrm{cat}_{m+1} \right>\,.
\end{equation*}
Interestingly, while the former family shares the same RoM as that of the family of star cat states, $\left| \mathrm{cat}_m^{*} \right>$, it is not hard to understand (even without solving the optimization problem) that the latter has a lower RoM. To do so, one can easily show that the states of the family $\left| \varphi_m^{*} \right>$ can be generated from the state $\left| + \right>^{\otimes m}$ via unitaries that can be synthesized using only $m$ $T$ gates. Moreover, it is easy to verify that, for the values of $m$ in Table~\ref{tab: RoM of cat states}, $\mathcal{R}\left( \left| \varphi_{m+1}^{*} \right> \right) = \mathcal{R}\left( \left| \mathrm{cat}_{m}^{*} \right> \right) = \mathcal{R}\left( \left| \phi_{m}^{*} \right> \right)$; which we conjecture holds for every $m$.

\subsubsection{Entanglement}

Even though multipartite entanglement is known to be hard to quantify, in the particular case of the original cat states, $\vert \text{cat}_m \rangle ,$ it is easy to see that all the entanglement is provided by the underlying (balanced) GHZ structure, associated with the maximum values for most multipartite entanglement monotones. This is so because cat states are equivalent to GHZ states up to local unitary operations. Hence, using the monotone introduced in Sec.~\ref{subsec: Background - RT entanglement}, Eq.~\eqref{eq: ent monotone}, we know that $\mathcal{E}(\vert \text{cat}_m\rangle) = \mathcal{E}(\vert \text{GHZ}_m)\rangle = 1$ for all $m$.  

A more technically subtle matter  pertains to the entanglement of the cat states $\vert \text{cat}_m^*\rangle$. In creating these states by measuring one of the qubits of the original cat states, it may appear that we are destroying entanglement entirely. Nevertheless, that is actually not the case. We can see this in two different ways. First, directly from the fact that we can use these states to inject entangling unitaries (cf. Sec.~\ref{subsec: Results - constructing the cat gadget}). Secondly, by using the multipartite entanglement monotone $\mathcal{E}$ to quantify the entanglement present in these states. Using this monotone we can show a simple result.
\begin{lemma}\label{lemma: new cat ent}
$\mathcal{E}(\vert \mathrm{cat}^*_m \rangle )= 1/2$, for all integers $m\geq 2$, with $\mathcal{E}$ given by Eq.~\eqref{eq: ent monotone}.
\end{lemma}We prove this simple fact in Appendix~\ref{app: entanglement lemma}. Since the monotone is faithful, the fact that $\mathcal{E}(\vert \text{cat}^*_m \rangle) = 1/2$ for all $m$ implies that star cat states are entangled states with less entanglement than the GHZ state (as quantified by $\mathcal{E}$). It also means that $\vert \text{cat}_m^*\rangle$ are not within the same local unitarily equivalence class of the (balanced) GHZ states, since $\mathcal{E}$ is invariant in such classes. A simple way of understanding why this is so is to write  $\vert \text{cat}_m^*\rangle$ in terms of the $\{ \left| T \right>, \left| T^{\perp} \right> \}$ basis so that:
\begin{equation*}
    \left| \mathrm{cat}_m^{*} \right> = \frac{1}{2} \left( (1+e^{-i\pi/4}) \left| T \right>^{\otimes m} + (1-e^{-i\pi/4}) \left| T^{\perp} \right>^{\otimes m}\right).
\end{equation*}
This equation highlights that the states $\left| \mathrm{cat}_m^{*} \right>$ are unbalanced-weight GHZ-type states, that is, they consist of a version of the original cat states where the coefficients of the orthogonal states $\left| T \right>$ and $\left| T ^{\perp}\right>$ have different norm. This leads to less than maximal entanglement, as suitably witnessed by $\mathcal{E}$.

In the following, we will use the fact that the states $\vert \text{cat}_m^*\rangle $ are non-stabilizer and entangled to fault-tolerantly inject interesting $m$-qubit unitaries. Through this procedure, we measure the states (hence destroying the entanglement) but use adaptivity to inject the resource into quantum circuits, in the form of entangling unitaries.

\subsection{Constructing a cat-gadget family}\label{subsec: Results - constructing the cat gadget}

\begin{figure*}
    \centering
    \includegraphics[width=0.8\textwidth]{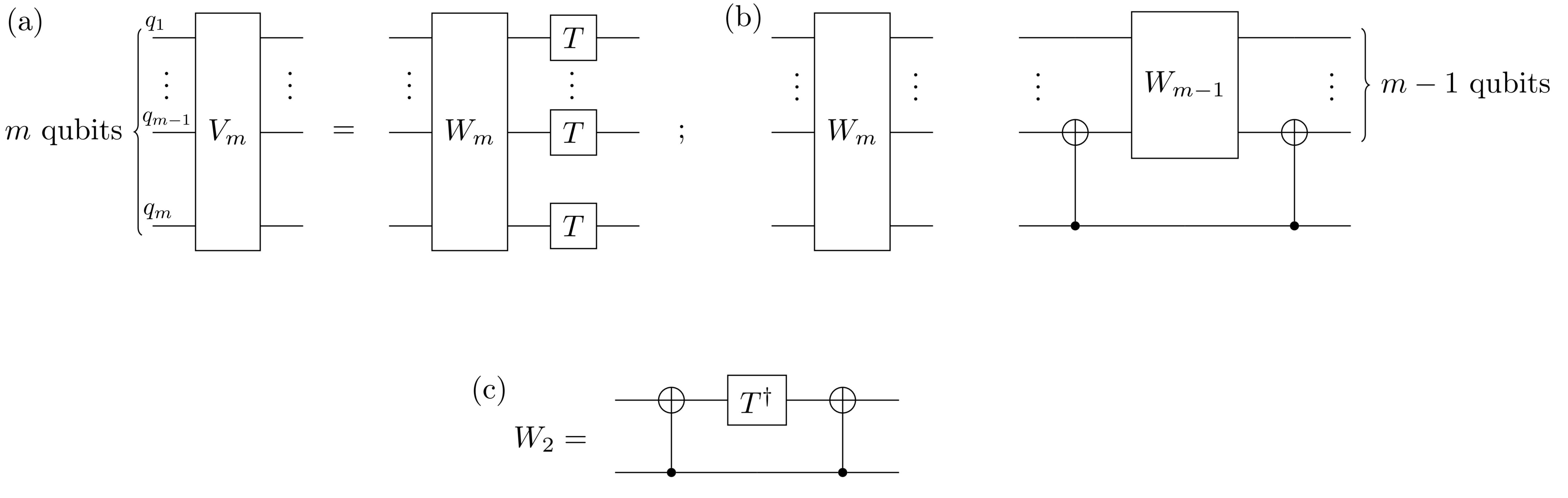}
    \caption{\textbf{Construction of the family of unitaries $V_m$.} In (a) we present the decomposition of $V_m$ into the entangling unitary $W_m$ followed by $T^{\otimes m}$; next, in (b), we show how $W_m$ can be constructed recursively from $W_2$, which is shown explicitly in (c).}
    \label{fig: V_m family construction}
\end{figure*}

\begin{figure*}[t]
    \centering
    \includegraphics[width=0.8\textwidth]{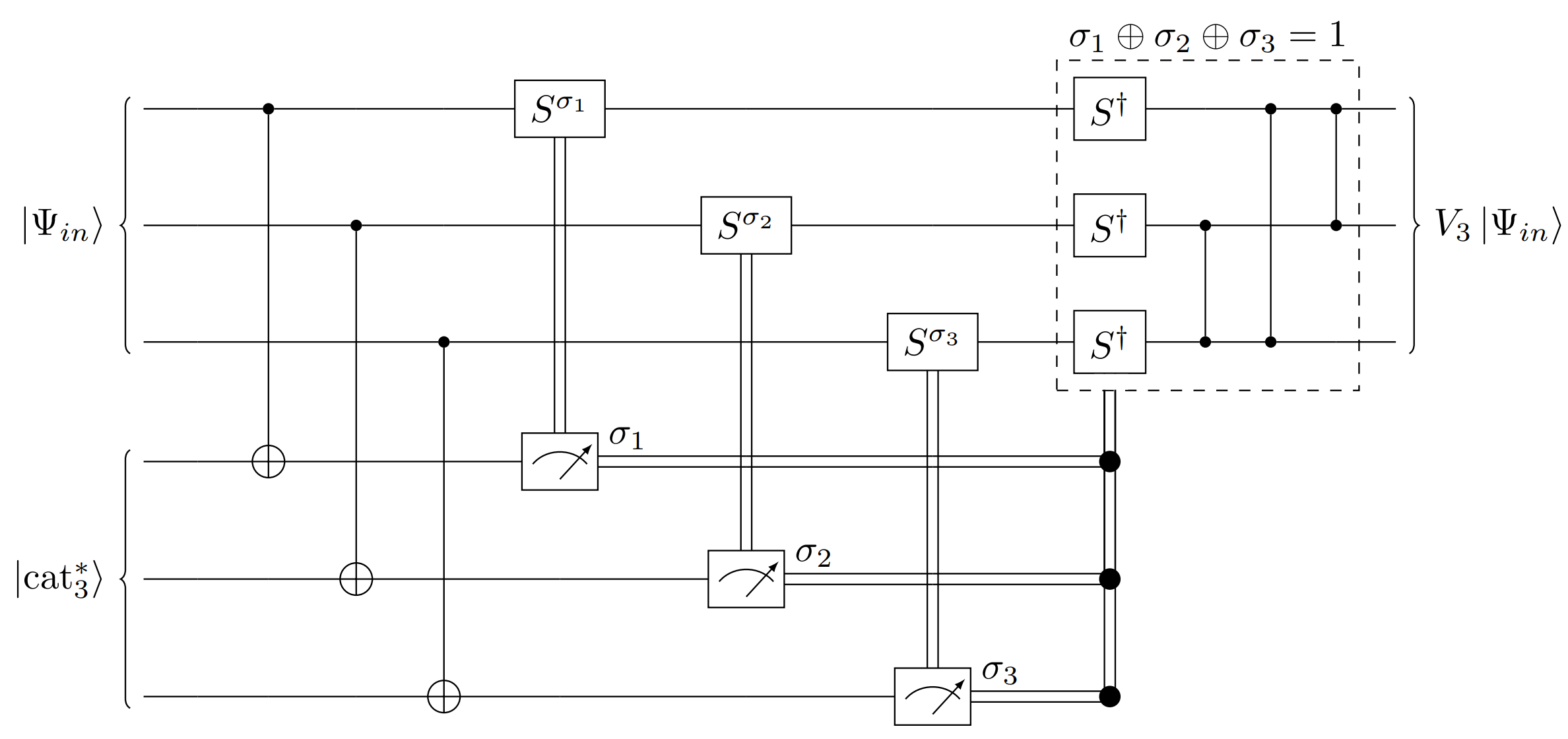}
    \caption{\textbf{Gadget that can be used to implement the unitary $V_3.$} The data qubits are entangled with the auxiliary qubits in the magic resource state $\left| \mathrm{cat}_3^{*} \right>$ using \textsc{cnot} gates controlled by the former and targeting the latter; this is then followed by computational basis measurements of the auxiliary qubits. After the appropriate Clifford corrections have been implemented (cf. Eq.~\eqref{eq: Clifford corrections}), the unitary transformation $V_3$,  from the family of unitaries $V_m:\text{ }V_m \left| + \right>^{\otimes m} = \left| \mathrm{cat}_m^{*} \right>,$ has been applied to the computational qubits.}
    \label{fig: Cat gadget for new cat 3}
\end{figure*}

As previously explained, we denote by $V_m$ the family of unitaries that generates the family of star cat states from the state $\left| +\right>^{\otimes m}$. These unitaries can be defined recursively with the aid of another family of unitaries $W_{m}$:
\begin{equation*}
    W_{m} = CX_{m,m-1} W_{m-1} CX_{m,m-1}
\end{equation*}
with $W_2 = CX_{2,1} T^{\dagger}_1 CX_{2,1}\,.$
Then,
\begin{equation*}
    V_{m} = T^{\otimes m}W_{m}\,.
\end{equation*}
For a visualization of the structure of these unitaries, see Fig.~\ref{fig: V_m family construction} (cf.~Fig.~4 of \cite{BeveCHK2019}); it is not hard to show that they belong to the third level of the Clifford hierarchy as they conjugate each $Z_i$ onto itself and each $X_i$ onto a Clifford unitary. Hence, Theorem~\ref{Theorem: Fault tolerance 3rd level}  guarantees that the star cat states can be created fault-tolerantly using only gates and measurements from the Clifford group. This highlights the existence of an alternative (and perhaps more convenient) preparation procedure for these states other than the one proposed in Eq.~\eqref{eq: new family of cat states}.  Assuming access to a factory of such magic resource states, we can (fault-tolerantly) implement any unitary $V_m$ via a \textit{cat gadget} which consumes the magic resource state $\left| \mathrm{cat}^{*}_m \right>$ but otherwise uses only Clifford operations and classical feedforward. To do so, we replace every $V_m$ in the quantum circuit by a sequence of \textsc{cnot} gates controlled in each of the data qubits $V_{m}$ acted on and targetting each of the qubits of the resource state $\left| \mathrm{cat}^{*}_m \right>\,;$ the auxiliary qubits are then measured in the computational basis. If all measurement outcomes, $\sigma_i$, yield 0, this immediately implements the desired unitary transformation $V_m$ on the computational qubits. Otherwise, some Clifford corrections are needed which take the form:
\begin{equation}
    C_m =
    \begin{cases}
        \bigotimes_{i=1}^{m} S^{\sigma_i}\,, & \text{if } \sigma^{\oplus} = 0\\
        \bigotimes_{i=1}^{m} \left( S^{\dagger} \right)^{1 - \sigma_i} \prod_{j=1}^{m-1} \prod_{k=j+1}^{m} CZ_{j,k}\,, & \text{otherwise}
    \end{cases}
    \label{eq: Clifford corrections}
\end{equation}
where we used $\sigma^{\oplus} = \bigoplus_{i=1}^{m} \sigma_i$ for brevity, with $\oplus$ representing addition modulo 2.

For a clearer visualization, Fig.~\ref{fig: Cat gadget for new cat 3} shows an example of the gadget that injects $V_3$. See Appendix~\ref{app: cat*_2 proof} for an example that illustrates how such gadgets can be found rather easily using the ZX-calculus framework. That framework can be used as formal proof that the gadget works for any $V_m$. Interestingly,  alternative (yet somewhat similar)  gadgets can be found that use the original cat states to inject these same unitaries.

It is important to guarantee that the number of gates associated with the Clifford corrections scales efficiently (i.e. polynomially) with the number of qubits. It is clear that that is indeed the case; the structure of the $CZ$ gates in the gadget corresponds to that needed to create a fully-connected graph, which has a total number of gates quadratic in $m$: $N_{CZ}=m(m-1)/2\,.$ This observation will be relevant below when discussing quantum resource reduction.

\subsubsection{Different structures}

The structure of the unitaries $V_m$, as depicted in Fig.~\ref{fig: V_m family construction}, is not unique. In fact, these unitaries can be implemented in a quantum circuit in many different ways. For instance, it is not hard to show that any of the $T$ gates appearing in the final layer after $W_m$ can be moved to stand before it. Additionally, because $W_2$ acts symmetrically on any pair of qubits, it can be flipped within the cascade of \textsc{cnot} gates a number of times given by $(m-1)$.

Another important observation is that the unitaries with $T$ inside the cascade of \textsc{cnot} gates (rather than $T^{\dagger}$) can also be implemented by injecting cat states with only minor changes to the Clifford corrections. 
The $T$ gate acting on the qubit which holds only the control for two \textsc{cnot} gates can also be pushed to be in between those two gates.

This (large) flexibility in the representation of these unitaries looks promising as it increases the chances that such structures can be found inside a given quantum circuit. In particular, it can be shown that the number of possible structures that can be injected by the cat gadget is (at least):
\begin{equation*}
    N_m = 3\times 2^{2m+1} \left(3m-5\right)\,.
\end{equation*}
We note that $N_m$ is lower bounded by $\Omega \left( 4^{m} m \right)$ and this exponential behavior to the number of structures gives us some hope that finding them inside quantum circuits is not as unlikely as one could think \textit{a priori}. Unfortunately, as we shall discuss in more depth in Sec.~\ref{subsec: Results - Applications}, this expectation did not materialize in the context of the quantum circuits studied herein.

\subsection{Applications}\label{subsec: Results - Applications}

\subsubsection{Classical simulation assisted by template matching}

\begin{table}[]
    \centering
    \begin{tabular}{p{2cm} >{\centering\arraybackslash}p{0.7cm} >{\centering\arraybackslash}p{0.7cm} >{\centering\arraybackslash}p{2cm}}
    \hline\hline
       Circuit  &  $n$  &  $t$  &  $V_2$ count \\
    \hline
        $\mathbf{adder_8}$              &  24 & 173 & \textbf{16}\\
        Adder8                          &  23 &  56 & 0 \\
        Adder16                         &  47 & 120 & 0 \\
        Adder32                         &  95 & 248 & 0 \\
        barenco tof3                    &   5 &  16 & 0 \\
        barenco tof4                    &   7 &  28 & 0 \\
        barenco tof5                    &   9 &  40 & 0 \\
        barenco tof10                   &  19 & 100 & 0 \\
        $\text{tof}_3$                  &   5 &  15 & 0 \\
        $\text{tof}_4$                  &   7 &  23 & 0 \\
        $\text{tof}_5$                  &   9 &  31 & 0 \\
        $\text{tof}_{10}$               &  19 &  71 & 0 \\
        $\text{csla-mux}_3$             &  15 &  62 & 0 \\
        $\text{csum-mux}_9$             &  30 &  84 & 0 \\
        $\text{gf(}2^4\text{)-mult}$    &  12 &  68 & 0 \\
        $\text{gf(}2^5\text{)-mult}$    &  15 & 115 & 0 \\
        $\text{gf(}2^6\text{)-mult}$    &  18 & 150 & 0 \\
        $\text{gf(}2^7\text{)-mult}$    &  21 & 217 & 0 \\
        \textbf{ham15-low}              &  17 &  97 & \textbf{1}\\
        ham15-med                       &  17 & 212 & 0 \\
        $\mathbf{hwb_6}$                &   7 &  75 & \textbf{1}\\
        mod-mult-55                     &   9 &  35 & 0 \\
        mod-red-21                      &  11 &  73 & 0 \\
        $\text{mod5}_4$                 &   5 &   8 & 0 \\
        $\textbf{nth-prime}\mathbf{_6}$ &   9 & 279 & \textbf{4}\\
        $\text{qcla-adder}_{10}$        &  36 & 162 & 0 \\
        $\text{qcla-com}_{7}$           &  24 &  95 & 0 \\
        $\text{qcla-mod}_{7}$           &  26 & 237 & 0 \\
        $\text{rc-adder}_{6}$           &  14 &  47 & 0 \\
        $\text{vbe-adder}_{3}$          &  10 &  24 & 0 \\
    \hline\hline
    \end{tabular}
    \caption{\textbf{Tested benchmark circuits.} We searched  for the patterns corresponding to the unitary $V_2$ in several of the benchmark circuits analyzed in Ref.~\cite{Kissinger2020reducing} and obtained from the optimization techniques proposed therein. The counts of $V_2$ detected by the pattern matching algorithm from Ref.~\cite{ItenMMSW2022} are presented in the last column. Rows presented in bold highlight the benchmark circuits for which such patterns were detected, otherwise, no such patterns were identified.}
    \label{tab: Benchmark circuits - list}
\end{table}
The fact that the original cat states are low rank (cf. Eq.~\eqref{eq: bounds on rank of cats}) led to the large improvement of the ZX simulator in Ref.~\cite{Kissinger+2022}. As previously explained, because an $m$-qubit cat state can be decomposed into a smaller number of stabilizer states than $\left| T \right>^{\otimes m},$ identifying ``cat-like'' structures in a given (graph-like) ZX-diagram allows us to decompose those structures into a smaller number of terms than we would have if we were trying to identify (and decompose) tensor products of $\left| T \right>$.

Naturally, in principle, the same will be true within the quantum circuit framework. Specifically, given a (general) unitary quantum circuit $U$ with $t$ $T$ gates, if we can identify any of the structures discussed in Sec.~\ref{subsec: Results - constructing the cat gadget}, then we may transform $U$ into an adaptive Clifford circuit whose resource state, $\left| \Psi_{R} \right>$, (potentially) involves copies of both $\left| T \right>$ and $\left| \mathrm{cat}^{*}_m \right>$ states.

To test this idea, we used the pattern-matching algorithm described in~\cite{ItenMMSW2022} and implemented in PennyLane~\cite{PennyLaneTM, PennyLane} to find the desired ``cat-like'' structures within the quantum circuits. We run the code on the benchmark circuits listed in Table~\ref{tab: Benchmark circuits - list} and also on the same family of exponentiated Pauli circuits as used in~\cite{Kissinger+2022} with the goal of finding $V_2$(-equivalent) structures. 

The results show that the improvements to classical simulation obtained by carrying out this procedure are either small or non-existent. In particular, we see that the pattern-matching algorithm detects structures that could be injected by $\left| \mathrm{cat}_2^{*} \right>$ only for four of the benchmark circuits. In the case of the exponentiated Pauli circuits, no such structures were found at all. This highlights the power of ZX-calculus. In that framework, it is easy to associate several phase gadgets with corresponding low-rank original cat states due to the reduction of the input circuit to a graph-like diagram via simplification rules. Contrarily, the comparative rigidity of quantum circuits does not allow the identification of comparable counts of structures associated with cat states.

To conclude this part, we should spare a comment on the cost of the pattern-matching algorithm proposed in~\cite{ItenMMSW2022} and used here. This is given by
\begin{equation*}
    \mathcal{O} \left( g_C^{g_P + 3} g_P^{g_P + 4} n_C^{n_P - 1}\right)\,,
\end{equation*}
where $n_C$ ($n_P$) and $g_C$ ($g_P$) are, respectively, the number of qubits and gates in the circuit (pattern). It is clear that while the algorithm is polynomial in the properties of the circuit, it is exponential in the properties of the pattern, $n_P$ and $g_P$. This limits our search capacity to structures of small (constant) size; otherwise, in trying to reduce the cost of classical simulation, we run into the risk of carrying out a search task that is more costly than the simulation itself. This search limitation is also not present in ZX-calculus, where the identification of phase gadgets is efficient~\cite{Kissinger+2022}.

As an example, we can consider the specific case of $V_2$ for which $n_P = 2$ and $g_P = 5$ so that: 
\begin{equation*}
    \mathcal{O} \left( 5^{9} g_C^{8} n_C \right)\,.
\end{equation*}
Note that $5^{9} = 1\,953\,125\,,$ which is a significantly large constant factor, and that the polynomial degree on the number of gates in the circuit is also quite large.

\subsubsection{Resource reduction in quantum computation}

Instead of focusing on classical simulation, we can consider the task of reducing the quantum resources needed for a certain quantum computation. In this case, doing pattern matching before running the quantum computer may allow us to identify unitary structures that can be injected via cat states $\left| \mathrm{cat}_m^{*} \right>.$ For each structure we detect in this way, we need one less (auxiliary) qubit to run the computation. For instance, consider the case of the benchmark circuit $\text{adder}_8$. This circuit has 24 qubits and 173 $T$ gates. As explained in Sec.~\ref{subsec: Background - gadgets}, fault-tolerance computations are performed such that each $T$ gate is injected using the $T$ gadget as long as we have access to a factory of magic states $\left| T \right>.$ This would require 173 auxiliary qubits in the state $\left| T \right>^{\otimes 173}.$ Alternatively, because 16 $V_2$-like structures are detected within the circuit, we could use only 157 auxiliary qubits in the state: $\left| \Psi_{R} \right> = \left| T \right>^{\otimes 125} \left| \mathrm{cat}_2^{*}\right>^{\otimes 16},$ so long as we have access to $\left| \mathrm{cat}_2^{*}\right>$ factories.\footnote{Note that in this example we are only analyzing things at a physical level, and disregard the added costs of encoding information into logical qubits comprised of several physical ones.}

Note that the same resource reduction would not be achieved if the distilled family of states were the original $\left| \text{cat}_m \right>$ state. This highlights the usefulness of protocols that ``compile'' magic from higher-dimensional states to lower-dimensional ones (without any loss of that resource).

\begin{figure*}[t]
    \centering
    \includegraphics[width=1\textwidth]{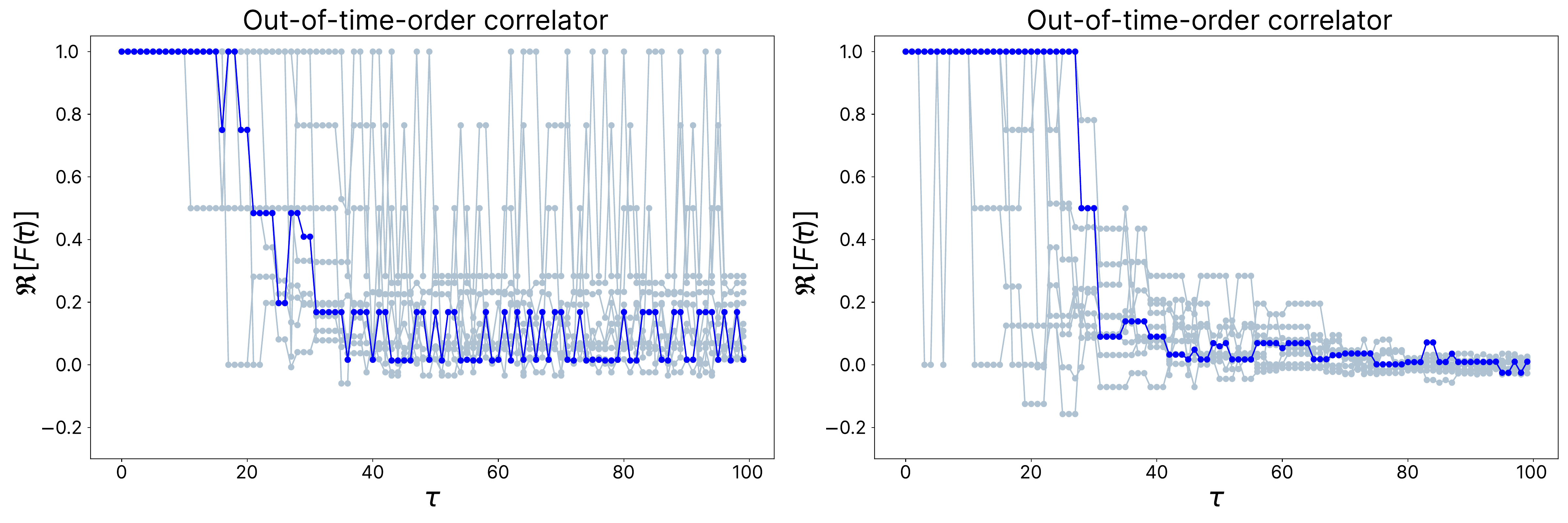}
    \caption{\textbf{Emergence of scrambling from cat-state injection.} As before, we simulate the OTOC given by Eq.~\eqref{eq: OTOC} considering the same evolution choices made in Fig.~\ref{fig: Ideal scrambling} other than the evolution of the unitary $U$. Here we have blocks of random non-entangling Clifford unitaries from the set $\{H,S\}$. At some time steps $\tau$, characterized by blocks of a fixed number of unitaries, we inject in random positions over the $10$-qubit circuit the cat-state unitary structures $V_m$. On the left (Fig.~\ref{fig: Cat Emegent Scrambling}(a)) we perform $16$ injections and on the right (Fig.~\ref{fig: Cat Emegent Scrambling}(b)) $36$.  The injection times are chosen arbitrarily and do not affect the results qualitatively. We inject unitaries $V_m$ via states $\{\vert \text{cat}^{*}_m \rangle\}_{m=2}^{10}$ uniformly at random, hence not necessarily $V_{10}$ but possibly unitaries having less magic and entanglement. Even though the circuits $V_m$ have a specific structure, they induce scrambling of quantum information in a controlled way, as the number of injected cat states increases.} 
    \label{fig: Cat Emegent Scrambling}
\end{figure*}

Besides the number of qubits, other features of quantum circuits, such as the depth or the number of two-qubit gates, are also deserving of attention when it comes to the optimization of quantum resources. Reducing the depth is relevant since it lowers the demands on the coherence time of the qubits, whilst minimizing the number of two-qubit gates is important because these gates tend to have lower fidelity than the single-qubit ones. Let us now analyze the CNOT gate count of our injection procedure. We note that the unitaries $V_m$ require exactly $N_{\textsc{cnot}} = 2(m-1)$ \textsc{cnot} gates, and cannot be implemented with fewer \textsc{cnot}s ~\cite{Cowtan2020phasegadget}. Comparatively, the gadget that injects this unitary using the state $\left| \mathrm{cat}_m^* \right>$ requires, on average, $N_{\textsc{cnot}}^{\prime} = m(m+3)/4.$ This comes from taking into account the $m$ \textsc{cnot} gates used to connect the computational qubits to the auxiliary qubits in the state $\left| \mathrm{cat}_m^* \right>$ (see Fig.~\ref{fig: Cat gadget for new cat 3}) plus the average number of $CZ$ gates in the Clifford corrections which is given by $m(m-1)/4$. It is not hard to see that $N_{\textsc{cnot}}^{\prime} > N_{\textsc{cnot}}$ $\forall m \geq 2.$ This means that the fault-tolerant implementation provided by the gadget actually requires more \textsc{cnot} gates than the direct implementation of the unitary itself.

This conclusion is different for applications where the non-local Clifford corrections in the dashed box of Fig.~\ref{fig: Cat gadget for new cat 3} might be irrelevant and, thus, dispensed with. Foregoing the application of these non-local corrections corresponds to applying either $V_m$ if the measurement outcomes are such that $\bigoplus_{i=1}^m \sigma_i = 0,$ or the unitary with $T$ inside the \textsc{cnot} cascade (rather than $T^{\dagger}$) if $\bigoplus_{i=1}^m \sigma_i = 1\,.$ In applications where this difference is irrelevant, such as, for instance, the study of scrambling in random quantum circuits discussed next, we can apply the gadget without the non-local correction leading to an implementation requiring only $N_{\textsc{cnot}}^{\prime} = m$ \textsc{cnot} gates, which is favorable over the number of such gates needed to directly implement $V_m$.

The latter setting is likely to be the exception rather than the rule since most interesting applications involve deterministic (rather than random) procedures, in which case the implementation of the proper Clifford corrections is vital. Therefore, we note that while this approach might reduce the number of qubits needed for a given quantum computation, we are likely incurring an increase in the number of controlled-\textsc{not} gates needed.

\subsubsection{Scrambling and quantum chaos of cat-doped random circuits}

We may use the family of star cat states to fault-tolerantly inject sub-circuit structures that induce scrambling of quantum information into a given circuit. In particular, it is interesting to see this kind of dynamic phenomenon emerge in circuits that lack both magic and entanglement, as both properties are injected with states from this family.

In Fig.~\ref{fig: Cat Emegent Scrambling} we show that scrambling of quantum information is induced by injection of cat-state unitaries $V_m$ with $2\leq m\leq 10$ into random quantum circuits. We allow the injection of $V_m$ with randomly chosen $m$. Assuming a circuit platform that has free access to Clifford gates drawn from the gate set $\{H,S\}$, we inject unitaries $V_m$ at random time-steps $\tau$, acting on at most $10$ qubits.

The plot in Fig.~\ref{fig: Cat Emegent Scrambling}(a) shows the behavior of the real part of the OTOC for 16 injections happening at random times up to step  $\tau=40$. The specifics of the unitary dynamics are exactly the same as in Fig.~\ref{fig: Ideal scrambling}, the only difference being the unitary blocks $U$ and the injection of $V_m$ unitaries. We see that after we stop injecting cat-states into the separable circuit the OTOC  stabilizes. The plot of Fig.~\ref{fig: Cat Emegent Scrambling}(b)   corresponds to  36 injections of cat-states and we see that after all injections have happened, up to time-step $\tau=90$ in this case, we have the generic feature of scrambling dynamics that was also present in universal Clifford+$T$ circuits (cf. Fig.~\ref{fig: Ideal scrambling}(c)). In particular, we note that cat injection allows for a controlled introduction of scrambling into these separable random circuits.

For a small number of injections and these kinds of circuits, it is possible for the OTOC to stabilize in non-zero values, as can be seen by the highlighted OTOC dynamics in Fig.~\ref{fig: Cat Emegent Scrambling}(a). This implies that, afterward, the OTOC should not decrease towards zero -- the intuition is that there is no more magic or entanglement to allow for that behavior -- and any such decrease would need to arise from decoherence acting on the system's evolution. The problem of distinguishing between decoherence and scrambling is of great interest within the community~\cite{Touil2021,Yoshida2019,Harris2022}.
The kind of slow and controlled scrambling induced by cat states may be useful in the characterization of the different quantum resources at play, in particular, to distinguish decoherence dynamics from the stable evolution of the OTOC after a finite number of injections. 

Known theoretically obtained results have shown that $T$-doped Clifford circuits require $O(n^2)$ non-Clifford gates, where $n$ is the number of qubits, to induce chaotic dynamics that is both hard to simulate and learn~\cite{Leone2021quantumchaos}. In our numerical simulations, this means that $\mathcal{O}(n)$ cat-injections -- for which $\mathcal{O}(n^2)$ $T$ gates and $\mathcal{O}(n)$ $CX$ gates are injected -- should suffice for achieving a dynamics having an exponential decay without recovery of the OTOC~\cite{Leone2022learning}.

If we consider random unitaries built from a universal gate set such as $\{H,CX,T\}$, the OTOC ceases to signal scrambling of information when any gate from the universal gate set is removed. Much is discussed in the literature regarding the necessary number of $T$ gates or $CX$ gates to introduce this kind of chaotic evolution, but recent results in classical simulation of quantum systems~\cite{aharonov2022polynomialtime} point towards revisiting the idea of classical simulability using sums over Feynman paths, where $H$ gates lead to an exponential overhead in classical simulation.

Finally, it is useful to point out that although the real part of the OTOC may suffer exponential decay in Clifford dynamics, the \textit{imaginary} part of the OTOC does not change. As a matter of fact, neither Clifford, nor incoherent or separable dynamics are able to change the imaginary part of the OTOC. The latter is a normally underappreciated quantity, but, in this case, $\mathfrak{I}[F(\tau)] \neq 0$ signals the presence of all elements in the gate set $\{H,CX,T\}$ in generic quantum circuits.  

It is interesting to point out that the imaginary part of the OTOC is not a representative witness of universality in general. For instance, the gate set $\{H, \text{Toffoli}\}$ is universal for quantum computation  but it also has the property that $\mathfrak{I}[F(\tau)] = 0$. Even still, the imaginarity of the OTOC may find fruitful applications elsewhere; for instance, it serves as a witness of basis-independent coherence~\cite{Wagner2023circuits}, and it has recently been applied as a tool in quantum machine learning~\cite{manas2023imaginary}.

\section{Discussion}\label{sec: Discussion}

In this work, we assessed the usefulness of cat states within the circuit model of quantum computation. To that end and assuming access to a factory of such non-stabilizer and entangled resource states, we constructed a family of gadgets that fault-tolerantly inject interesting unitaries into quantum circuits. Then, we explored these gadgets in the context of three main applications.

First, we considered the task of (strong) classical simulation of quantum circuits. Specifically, inspired by similar work in the context of the ZX-calculus framework, we explored the idea of searching for specific (non-Clifford) cat-state-like unitaries within benchmark circuits and exponentiated Pauli circuits. The idea is to use the identification of such structures to improve the cost of classical simulation, as they can be implemented via the injection of low-rank resource states. Interestingly, our results show that while cat-state-like structures may appear commonly in ZX-calculus after simplifications have taken place, this does not happen in general circuits due to their (more) rigid structure. This suggests that the success of strong simulation using cat states within the ZX-calculus paradigm arises from simplifications made possible by the diagrammatic representations. 

Secondly, we noticed that the application of these gadgets can also be exploited in the context of resource minimization in actual quantum computation. 
We explained how the detection of cat-like structures within quantum circuits can lead to a reduction in the number of qubits needed to perform the corresponding (fault-tolerant) computation in actual quantum hardware. This is a clear advantage over the ZX-calculus framework, wherein the simplifications (whilst leading to faster classical simulation) come at the cost of departing from the quantum circuit model. We also think that the unitary structures injected by our gadgets may find applications in several protocols. For instance,  structures akin to these appear in the context of quantum parameter estimation. 

It is also interesting to remark that whilst the gadget used to inject the non-stabilizer and entangling unitaries $V_m$ does not use any magic gates, it does require the use of two-qubit entangling gates. This observation highlights a crucial difference between the two resources. Moreover, in Sec.~\ref{subsec: Results - Applications}, we discussed that our gadget requires a larger number of controlled-\textsc{not} gates to inject $V_m$ than the one needed to implement $V_m$ directly. This is in contrast with what happens, for instance, with the gadget that injects the Toffoli gate which requires 4.5 and 6 \textsc{cnot} gates, respectively on average and in the worst case. On the other hand, the optimal decomposition for the Toffoli gate uses 6 \textsc{cnot}s \cite{ShendeMarkov2009} which means that the gadget is advantageous in terms of the count of two-qubit entangling gates. This leads to two questions: first, whether the cat gadget is improvable in terms of the number of \textsc{cnot}s needed, and second, which are the entangling unitaries that might be injected with an advantage in the number of \textsc{cnot}s. 

Finally, we moved towards understanding how the evident nonclassical resources of cat states can be used to induce scrambling of quantum information into otherwise non-entangling circuits. We showed that these states can be used to generate scrambling of information in a slow and controlled way.
We have seen numerically that injection of a sufficient number of cat states can induce dynamics that should be hard to simulate classically. In particular, this might have applications in the quantification of different nonclassical resources in quantum information scrambling.

\begin{acknowledgments}
FCRP and RW contributed equally to this work. The authors would like to thank Mark Howard for comments on the text, and also for calling attention to Ref. \cite{BeveCHK2019} and some similarities with the work therein. FCRP and RW acknowledge support from FCT -- Fundação para a Ciência e a Tecnologia (Portugal) through PhD Research Scholarship 2020.07245.BD and PhD Grant SFRH/BD/151199/2021, respectively. EFG acknowledges support from FCT –
Fundação para a Ciência e a Tecnologia (Portugal) via project CEECINST/00062/2018. This work was supported by the Digital Horizon Europe project FoQaCiA, GA no.101070558.
\end{acknowledgments}

\bibliography{bibliography}

\appendix
\onecolumngrid
\section{Basics of ZX-calculus}\label{app: ZX-calculus basics}

In this Appendix, we review the basics of the ZX-calculus. For our purposes, this calculus can be seen simply as a (formal) language allowing us to perform calculations using operations over diagrams as opposed to the commonly used matrix calculus. In particular, we shall use diagrams to perform calculations in order to discover quantum circuits that can be used for the specific task of injecting cat states. It is worth noting that this is just one of many aspects associated with research in the ZX-calculus, and we refer to Refs.~\cite{Coecke2011interacting,coecke2017picturing,wetering2020zxcalculus} for deeper discussions. Similar in spirit to Feynmann's ubiquitous diagrams in high energy physics, we will draw parallels between the gates necessary to describe quantum circuits as diagrams, to later prescribe \textit{rules} that are only diagrammatic but that inform us how to reason using the diagrams only, forgetting the matrices until it is convenient.

States, measurement effects, and quantum gates have ZX counterparts. The building blocks of this diagrammatic language are the so-called \textit{spiders} that correspond to green or red nodes having various lines as input and/or output.

Green spiders correspond to $Z$ rotations while red spiders correspond to $X$ rotations.\newline 
\begin{center}
\begin{tikzpicture}
	\begin{pgfonlayer}{nodelayer}
		\node [style=gn] (0) at (0, 0) {$\alpha$};
		\node [style=none] (1) at (-1, 1) {};
		\node [style=none] (2) at (1, 1) {};
		\node [style=none] (3) at (0, 0.75) {$\dots$};
		\node [style=none] (4) at (-1, -1) {};
		\node [style=none] (5) at (1, -1) {};
		\node [style=none] (6) at (0, -1) {$\dots$};
		\node [style=none] (7) at (1.25, 0) {$=$};
		\node [style=none] (8) at (0, 1.25) {$n$ times};
		\node [style=none] (9) at (4.5, 0) {$\vert\underbrace{ 0 \dots 0 }_{n}\rangle \langle 0 \dots 0 \vert + e^{i\alpha}\vert 1 \dots 1 \rangle \langle 1 \dots 1 \vert$};
	\end{pgfonlayer}
	\begin{pgfonlayer}{edgelayer}
		\draw [bend left=15] (0) to (1.center);
		\draw [bend right=15] (0) to (2.center);
		\draw [bend left=15] (0) to (5.center);
		\draw [bend right=15] (0) to (4.center);
	\end{pgfonlayer}
\end{tikzpicture}    
\end{center}

\begin{center}
\scalebox{0.88}{
\begin{tikzpicture}
	\begin{pgfonlayer}{nodelayer}
		\node [style=none] (1) at (-1, 1) {};
		\node [style=none] (2) at (1, 1) {};
		\node [style=none] (3) at (0, 0.75) {$\dots$};
		\node [style=none] (4) at (-1, -1) {};
		\node [style=none] (5) at (1, -1) {};
		\node [style=none] (6) at (0, -1) {$\dots$};
		\node [style=none] (7) at (1.25, 0) {$=$};a
		\node [style=none] (8) at (0, 1.25) {$n$ times};
		\node [style=none] (9) at (5, 0) {$\vert\underbrace{ + \dots + }_{n}\rangle \langle + \dots + \vert + e^{i\alpha}\vert - \dots - \rangle \langle - \dots - \vert$};
		\node [style=rn] (10) at (0, 0) {$\alpha$};
	\end{pgfonlayer}
	\begin{pgfonlayer}{edgelayer}
		\draw [bend right=15] (1.center) to (10);
		\draw [bend right=15] (10) to (2.center);
		\draw [bend left=15] (10) to (5.center);
		\draw [bend right=15] (10) to (4.center);
	\end{pgfonlayer}
\end{tikzpicture}
}    
\end{center}

This definition allows us to represent various states, quantum gates, and measurements diagrammatically. Some of the diagrams we will use are shown below; we note that all diagrams are represented up to scalar factors. \newline 

\begin{center}
    \scalebox{0.9}{
\begin{tikzpicture}
	\begin{pgfonlayer}{nodelayer}
		\node [style=rn] (0) at (0, 0) {};
		\node [style=none] (1) at (1, 1) {};
		\node [style=none] (2) at (1, 0) {};
		\node [style=none] (3) at (1, -1) {};
		\node [style=none] (4) at (-2, 0) {$\vert \mathrm{GHZ}_3 \rangle$};
		\node [style=none] (5) at (-1, 0) {$=$};
		\node [style=gn] (6) at (4.5, 1) {};
		\node [style=none] (7) at (2, 0) {$CX$};
		\node [style=none] (8) at (3, 0) {$=$};
		\node [style=rn] (9) at (4.5, -1) {};
		\node [style=none] (10) at (3.5, 1) {};
		\node [style=none] (11) at (5.5, 1) {};
		\node [style=none] (12) at (3.5, -1) {};
		\node [style=none] (13) at (5.5, -1) {};
		\node [style=gn] (14) at (0, -2) {$\frac{\pi}{4}$};
		\node [style=none] (15) at (-1, -2) {};
		\node [style=none] (16) at (1, -2) {};
		\node [style=none] (17) at (-2, -2) {$=$};
		\node [style=none] (18) at (-3, -2) {$T$};
		\node [style=none] (19) at (2, -2) {$S$};
		\node [style=none] (20) at (3, -2) {$=$};
		\node [style=gn] (21) at (4.5, -2) {$\frac{\pi}{2}$};
		\node [style=none] (22) at (3.5, -2) {};
		\node [style=none] (23) at (5.5, -2) {};
		\node [style=none] (24) at (-0.75, -3) {$H$};
		\node [style=none] (25) at (0.25, -3) {$=$};
		\node [style=none] (26) at (1.25, -3) {};
		\node [style=none] (27) at (3.25, -3) {};
		\node [style=hadamard] (28) at (2.25, -3) {};
	\end{pgfonlayer}
	\begin{pgfonlayer}{edgelayer}
		\draw [bend left] (0) to (1.center);
		\draw (0) to (2.center);
		\draw [bend right] (0) to (3.center);
		\draw (6) to (9);
		\draw (12.center) to (13.center);
		\draw (10.center) to (11.center);
		\draw (15.center) to (16.center);
		\draw (22.center) to (23.center);
		\draw (27.center) to (26.center);
	\end{pgfonlayer}
\end{tikzpicture}
}
\end{center}

Above, $\vert \mathrm{GHZ}_3 \rangle = \frac{1}{\sqrt{2}}(\vert \mathrm{+++}\rangle + \vert \mathrm{---}\rangle)$ stands for the Greenberger-Horne-Zeilinger (GHZ) state~\cite{greenberger1989going}. To represent measurements, we describe nodes using the notation $\sigma\pi$ with $\sigma \in \{0,1\}$, where \begin{tikzpicture}
	\begin{pgfonlayer}{nodelayer}
		\node [style=rn] (0) at (0, 0) {$\sigma\pi$};
		\node [style=none] (1) at (-2, 0) {};
	\end{pgfonlayer}
	\begin{pgfonlayer}{edgelayer}
		\draw (1.center) to (0);
	\end{pgfonlayer}
\end{tikzpicture} represents a $Z$ measurement with outcome $\sigma$ and  \begin{tikzpicture}
	\begin{pgfonlayer}{nodelayer}
		\node [style=gn] (0) at (0, 0) {$\sigma\pi$};
		\node [style=none] (1) at (-2, 0) {};
	\end{pgfonlayer}
	\begin{pgfonlayer}{edgelayer}
		\draw (1.center) to (0);
	\end{pgfonlayer}
\end{tikzpicture}
 represents an $X$ measurement with outcome $\sigma$. This helps us to cope with the uncertainty of measurement in quantum theory.

To perform calculations, one must describe a set of allowed rules for the diagrammatic language. For our purposes, we need only a subset of the complete identities of ZX-calculus. A set of rules that is sufficient for proving universality and completeness can be found in Refs.~\cite{jeandel2020completeness}. We will constantly use the following rules: ($f$) fusion of spiders, where spiders with the same color can be merged into a single spider by adding the labels of the original spiders.
\begin{center}
    \scalebox{0.9}{
\begin{tikzpicture}
	\begin{pgfonlayer}{nodelayer}
		\node [style=gn] (0) at (-1, 1) {$\alpha$};
		\node [style=gn] (1) at (1, -1) {$\beta$};
		\node [style=gn] (2) at (4, 0) {$\alpha + \beta$};
		\node [style=none] (3) at (0, 2) {};
		\node [style=none] (4) at (-2, 2) {};
		\node [style=none] (5) at (0, -2) {};
		\node [style=none] (6) at (2, -2) {};
		\node [style=none] (7) at (0, 0) {$\dots$};
		\node [style=none] (8) at (-1, 1.75) {$\dots$};
		\node [style=none] (9) at (1, -1.75) {$\dots$};
		\node [style=none] (10) at (2, 0) {$\stackrel{(f)}{=}$};
		\node [style=none] (11) at (3, 1) {};
		\node [style=none] (12) at (5, 1) {};
		\node [style=none] (13) at (3, -1) {};
		\node [style=none] (14) at (5, -1) {};
	\end{pgfonlayer}
	\begin{pgfonlayer}{edgelayer}
		\draw [bend left=15] (5.center) to (1);
		\draw [bend right=15] (6.center) to (1);
		\draw [bend left=15, looseness=0.75] (0) to (4.center);
		\draw [bend right=15] (0) to (3.center);
		\draw [bend left=45, looseness=1.25] (1) to (0);
		\draw [bend left] (0) to (1);
		\draw [bend left=15] (13.center) to (2);
		\draw [bend left=15] (2) to (14.center);
		\draw [bend left=15] (2) to (11.center);
		\draw [bend right=15] (2) to (12.center);
	\end{pgfonlayer}
\end{tikzpicture}

}
\end{center}
Another important rule is the copy rule, that we refer to as $(\pi)$. This rule will be particularly relevant for us whenever we want to study Clifford corrections and propagate the results of past outcomes through the diagram. Single-system spiders with angles that are multiples of $\pi$ can be copied through generic-label spiders of the opposite color.

\begin{center}
    \scalebox{0.9}{
\begin{tikzpicture}
	\begin{pgfonlayer}{nodelayer}
		\node [style=rn] (0) at (-3, 0) {$\sigma\pi$};
		\node [style=gn] (1) at (-2, 0) {$\alpha$};
		\node [style=none] (2) at (-1, 1) {};
		\node [style=none] (3) at (-1, 0) {};
		\node [style=none] (4) at (-1, -1) {};
		\node [style=none] (5) at (-4, 0) {};
		\node [style=none] (6) at (0, 0) {$\stackrel{(\pi)}{=}$};
		\node [style=rn] (7) at (3.25, 1) {$\sigma\pi$};
		\node [style=gn] (8) at (2, 0) {$(-1)^\sigma\alpha$};
		\node [style=none] (9) at (4, 1) {};
		\node [style=none] (10) at (4, 0) {};
		\node [style=none] (11) at (4, -1) {};
		\node [style=none] (12) at (1, 0) {};
		\node [style=rn] (13) at (3.25, 0) {$\sigma\pi$};
		\node [style=rn] (14) at (3.25, -1) {$\sigma\pi$};
	\end{pgfonlayer}
	\begin{pgfonlayer}{edgelayer}
		\draw (5.center) to (1);
		\draw [bend left] (1) to (2.center);
		\draw (1) to (3.center);
		\draw [bend right] (1) to (4.center);
		\draw (12.center) to (8);
		\draw [bend left] (8) to (9.center);
		\draw (8) to (10.center);
		\draw [bend right] (8) to (11.center);
	\end{pgfonlayer}
\end{tikzpicture}
}
\end{center}

There are some identities that are useful as they simplify elements into the identity gate, that in the ZX-calculus is simply represented as a single wire.
\begin{center}
\begin{tikzpicture}
	\begin{pgfonlayer}{nodelayer}
		\node [style=none] (0) at (-2, 1) {};
		\node [style=none] (1) at (-2, -1) {};
		\node [style=none] (2) at (-1, 0) {$=$};
		\node [style=none] (3) at (0, 1) {};
		\node [style=none] (4) at (0, -1) {};
		\node [style=gn] (5) at (0, 0) {};
		\node [style=none] (6) at (1, 0) {$=$};
		\node [style=none] (7) at (2, 1) {};
		\node [style=none] (8) at (2, -1) {};
		\node [style=rn] (9) at (2, 0) {};
		\node [style=none] (10) at (3, 0) {$=$};
		\node [style=none] (11) at (4, 1) {};
		\node [style=none] (12) at (4, -1) {};
		\node [style=rn] (13) at (4, 0) {$2\pi$};
	\end{pgfonlayer}
	\begin{pgfonlayer}{edgelayer}
		\draw (1.center) to (0.center);
		\draw (4.center) to (3.center);
		\draw (8.center) to (7.center);
		\draw (12.center) to (11.center);
	\end{pgfonlayer}
\end{tikzpicture}
\end{center}

The last identity we consider is the phase-gadget identity~\cite{Cowtan2020phasegadget,Kissinger2020reducing} which teaches us how to represent a ZX-calculus phase gadget into a diagram resembling a quantum circuit. It works for an arbitrary number of qubits, $n$, but for simplicity we make a finite representation.

\begin{center}
    \scalebox{0.9}{
\begin{tikzpicture}
	\begin{pgfonlayer}{nodelayer}
		\node [style=none] (0) at (-5, 0) {};
		\node [style=none] (1) at (-2.75, 0) {};
		\node [style=none] (2) at (-5, -1) {};
		\node [style=none] (3) at (-2.75, -1) {};
		\node [style=none] (4) at (-5, -2) {};
		\node [style=none] (5) at (-2.75, -2) {};
		\node [style=none] (6) at (-2.75, -3) {};
		\node [style=none] (7) at (-5, -3) {};
		\node [style=gn] (8) at (-4, 0) {};
		\node [style=gn] (9) at (-4, -1) {};
		\node [style=gn] (10) at (-4, -2) {};
		\node [style=gn] (11) at (-4, -3) {};
		\node [style=rn] (12) at (-3, 1) {};
		\node [style=gn] (13) at (-3, 2) {$\alpha$};
		\node [style=none] (14) at (-2.25, -1.5) {$=$};
		\node [style=none] (15) at (-1.75, 0) {};
		\node [style=none] (16) at (4.25, 0) {};
		\node [style=none] (17) at (-1.75, -1) {};
		\node [style=none] (18) at (4.25, -1) {};
		\node [style=none] (19) at (-1.75, -2) {};
		\node [style=none] (20) at (4.25, -2) {};
		\node [style=none] (21) at (4.25, -3) {};
		\node [style=none] (22) at (-1.75, -3) {};
		\node [style=gn] (28) at (1.25, 0) {$\alpha$};
		\node [style=rn] (29) at (0.5, 0) {};
		\node [style=rn] (30) at (-0.25, -1) {};
		\node [style=rn] (31) at (-1, -2) {};
		\node [style=rn] (32) at (2, 0) {};
		\node [style=rn] (33) at (2.75, -1) {};
		\node [style=rn] (34) at (3.5, -2) {};
		\node [style=gn] (35) at (0.5, -1) {};
		\node [style=gn] (36) at (-0.25, -2) {};
		\node [style=gn] (37) at (-1, -3) {};
		\node [style=gn] (38) at (2, -1) {};
		\node [style=gn] (39) at (2.75, -2) {};
		\node [style=gn] (40) at (3.5, -3) {};
	\end{pgfonlayer}
	\begin{pgfonlayer}{edgelayer}
		\draw (0.center) to (1.center);
		\draw (2.center) to (3.center);
		\draw (4.center) to (5.center);
		\draw (7.center) to (6.center);
		\draw (8) to (12);
		\draw (12) to (9);
		\draw (12) to (10);
		\draw (12) to (11);
		\draw (12) to (13);
		\draw (15.center) to (16.center);
		\draw (17.center) to (18.center);
		\draw (19.center) to (20.center);
		\draw (22.center) to (21.center);
		\draw (29) to (35);
		\draw (36) to (30);
		\draw (37) to (31);
		\draw (32) to (38);
		\draw (33) to (39);
		\draw (34) to (40);
	\end{pgfonlayer}
\end{tikzpicture}

}
\end{center}

The ZX-calculus phase-gadget identity greatly simplifies the proof that our cat gadget works since we can view the injection of $\vert \text{cat}_m^*\rangle$ in the ZX-calculus as the injection of a specific phase-gadget into the circuit; the remaining complication is to find out what are the Clifford corrections to be performed and the complete list of equivalent unitaries that can be injected.

Using these tools, Refs.~\cite{Kissinger+2022,Kissinger2020reducing} studied strong simulation as discussed in the main text. In particular, as it is now clear, one can write any quantum circuit into a ZX-calculus form. Once we perform the simplifications, the final diagram can be put in what is called a graph-like ZX-diagram, that is, a diagram with only green spiders and Hadamard edges (i.e., edges that have a Hadamard gate on them). Thus, graph-like ZX-diagrams are node-weighted graphs~\footnote{Graphs can have weights, which means that they have labels either in their nodes or edges. An edge-weighted graph is a graph that has a function inducing labels over the edges. Similarly, one can define node-weighted graphs.} with node weights representing non-Clifford spiders.

\section{ZX-calculus proof for $\left| \mathrm{cat}_2^{*} \right>$}\label{app: cat*_2 proof}

Cat states $\vert \text{cat}_m \rangle$ as defined in the main text have an elegant diagrammatic representation~\cite{Kissinger+2022}.\newline 
\begin{center}
\begin{tikzpicture}
	\begin{pgfonlayer}{nodelayer}
		\node [style=rn] (0) at (0, 0) {};
		\node [style=gn] (1) at (2, 2) {$\frac{\pi}{4}$};
		\node [style=gn] (2) at (2, 1) {$\frac{\pi}{4}$};
		\node [style=gn] (4) at (2, 0) {$\frac{\pi}{4}$};
		\node [style=gn] (5) at (2, -1.5) {$\frac{\pi}{4}$};
		\node [style=none] (6) at (2, -0.75) {$\vdots$};
		\node [style=none] (7) at (3.25, 2) {};
		\node [style=none] (8) at (3.25, 1) {};
		\node [style=none] (9) at (3.25, 0) {};
		\node [style=none] (10) at (3.25, -1.5) {};
		\node [style=none] (11) at (-1.25, 0) {$\vert \text{cat}_m \rangle := $};
		\node [style=none] (12) at (2.75, -0.75) {$m$ times};
	\end{pgfonlayer}
	\begin{pgfonlayer}{edgelayer}
		\draw [bend left=45] (0) to (1);
		\draw [bend left] (0) to (2);
		\draw (0) to (4);
		\draw [bend right=45, looseness=0.75] (0) to (5);
		\draw (1) to (7.center);
		\draw (2) to (8.center);
		\draw (4) to (9.center);
		\draw (5) to (10.center);
	\end{pgfonlayer}
\end{tikzpicture}
\end{center}

They correspond to GHZ states followed by the (local) action of $T$ gates in each of the system's qubits. In our work, we start by defining a family of star cat states, $\vert \text{cat}_{m-1}^*\rangle$, obtained by making a $Y$ measurement in one of the arms of the original cat states. In the ZX-calculus representation, this family of states can be depicted as seen below.
\begin{center}
    \scalebox{0.9}{
\begin{tikzpicture}
	\begin{pgfonlayer}{nodelayer}
		\node [style=rn] (0) at (0, 0) {};
		\node [style=gn] (1) at (-1.5, 0) {$-\frac{\pi}{4}$};
		\node [style=gn] (2) at (1.5, 2) {$\frac{\pi}{4}$};
		\node [style=gn] (3) at (1.5, 1) {$\frac{\pi}{4}$};
		\node [style=gn] (4) at (1.5, 0) {$\frac{\pi}{4}$};
		\node [style=gn] (5) at (1.5, -2.25) {$\frac{\pi}{4}$};
		\node [style=none] (6) at (1.5, -1) {$\vdots$};
		\node [style=none] (7) at (2.75, -1) {$m-1$ times};
		\node [style=none] (8) at (2.25, 2) {};
		\node [style=none] (9) at (2.25, 1) {};
		\node [style=none] (10) at (2.25, 0) {};
		\node [style=none] (11) at (2.25, -2.25) {};
		\node [style=none] (12) at (-4, 0) {$\vert \text{cat}^*_{m-1}\rangle $};
		\node [style=none] (13) at (-3, 0) {$=$};
	\end{pgfonlayer}
	\begin{pgfonlayer}{edgelayer}
		\draw (1) to (0);
		\draw (0) to (4);
		\draw [bend left] (0) to (3);
		\draw [bend left] (0) to (2);
		\draw [bend right] (0) to (5);
		\draw (5) to (11.center);
		\draw (10.center) to (4);
		\draw (9.center) to (3);
		\draw (8.center) to (2);
	\end{pgfonlayer}
\end{tikzpicture}

}
\end{center}

The choice for defining these states is somewhat arbitrary, as we wanted the relation $\vert \text{cat}_2^*\rangle = \vert CS \rangle$. However, this is not strictly necessary and the usual cat states can also be implemented fault-tolerantly into quantum circuits. We will demonstrate using the ZX-calculus the injection of $\vert \text{cat}_2^* \rangle $ in what follows.

We apply \textsc{cnot} gates and measure the $\vert \text{cat}_2^* \rangle $ state, as depicted in the ZX-diagram below. Using the ZX rules, we then simplify the diagram propagating the outcome values throughout the circuit. \newline

\begin{center}
\scalebox{0.9}{
\begin{tikzpicture}
	\begin{pgfonlayer}{nodelayer}
		\node [style=gn] (37) at (-7.75, 2.25) {$\frac{\pi}{4}$};
		\node [style=gn] (38) at (-7, 1.25) {$\frac{\pi}{4}$};
		\node [style=none] (43) at (-9, 3.25) {};
		\node [style=none] (44) at (-4, 3.25) {};
		\node [style=none] (45) at (-9, 4.25) {};
		\node [style=none] (46) at (-4, 4.25) {};
		\node [style=gn] (47) at (-5.75, 4.25) {};
		\node [style=gn] (48) at (-4.75, 3.25) {};
		\node [style=rn] (55) at (-9, 1.25) {};
		\node [style=gn] (58) at (-8, 0.25) {$-\frac{\pi}{4}$};
		\node [style=none] (59) at (-2.5, 2.25) {$\stackrel{(f)}{=}$};
		\node [style=none] (107) at (5.25, 3.25) {};
		\node [style=none] (108) at (10.25, 3.25) {};
		\node [style=none] (109) at (5.25, 4.25) {};
		\node [style=none] (110) at (10.25, 4.25) {};
		\node [style=gn] (111) at (7, 4.25) {};
		\node [style=gn] (112) at (8.25, 3.25) {};
		\node [style=rn] (113) at (7.5, 0.75) {$(\sigma_1+\sigma_2)\pi$};
		\node [style=gn] (114) at (7, 2.75) {$(-1)^{\sigma_1}\frac{\pi}{4}$};
		\node [style=gn] (115) at (8.25, 1.75) {$(-1)^{\sigma_2}\frac{\pi}{4}$};
		\node [style=gn] (116) at (9.25, 0.25) {$-\frac{\pi}{4}$};
		\node [style=none] (119) at (-8.75, -2) {};
		\node [style=none] (120) at (-4, -2) {};
		\node [style=none] (121) at (-8.75, -1) {};
		\node [style=none] (122) at (-4, -1) {};
		\node [style=gn] (123) at (-8.25, -1) {};
		\node [style=gn] (124) at (-7.5, -2) {};
		\node [style=rn] (125) at (-8, -3.5) {$(\sigma_1+\sigma_2)\pi$};
		\node [style=gn] (126) at (-6.25, -1) {$(-1)^{\sigma_1}\frac{\pi}{4}$};
		\node [style=gn] (127) at (-6.25, -2) {$(-1)^{\sigma_2}\frac{\pi}{4}$};
		\node [style=gn] (128) at (-7.25, -4.75) {$-\frac{\pi}{4}$};
		\node [style=none] (129) at (4.25, 2.25) {$\stackrel{(\pi)}{=}$};
		\node [style=none] (132) at (-1.75, -2) {};
		\node [style=none] (133) at (8.25, -2) {};
		\node [style=none] (134) at (-1.75, -1) {};
		\node [style=none] (135) at (8.25, -1) {};
		\node [style=gn] (136) at (0.75, -1) {};
		\node [style=gn] (137) at (2.5, -2) {};
		\node [style=rn] (138) at (1.5, -3) {};
		\node [style=gn] (139) at (6, -1) {$(-1)^{\sigma_1}\frac{\pi}{4}$};
		\node [style=gn] (140) at (6, -2) {$(-1)^{\sigma_2}\frac{\pi}{4}$};
		\node [style=gn] (141) at (1, -4) {$-\frac{\pi}{4}$};
		\node [style=rn] (143) at (3.5, -1) {$(\sigma_1+\sigma_2)\pi$};
		\node [style=rn] (144) at (-0.75, -1) {$(\sigma_1+\sigma_2)\pi$};
		\node [style=rn] (145) at (-3.75, 2.25) {$\sigma_1\pi$};
		\node [style=rn] (146) at (-3.75, 1.25) {$\sigma_2\pi$};
		\node [style=rn] (147) at (-5.75, 2.25) {};
		\node [style=rn] (148) at (-4.75, 1.25) {};
		\node [style=gn] (149) at (-0.25, 2.25) {$\frac{\pi}{4}$};
		\node [style=gn] (150) at (0.5, 1.25) {$\frac{\pi}{4}$};
		\node [style=none] (151) at (-1.5, 3.25) {};
		\node [style=none] (152) at (3.5, 3.25) {};
		\node [style=none] (153) at (-1.5, 4.25) {};
		\node [style=none] (154) at (3.5, 4.25) {};
		\node [style=gn] (155) at (1.75, 4.25) {};
		\node [style=gn] (156) at (2.75, 3.25) {};
		\node [style=rn] (157) at (-1.5, 1.25) {};
		\node [style=gn] (158) at (-0.5, 0.25) {$-\frac{\pi}{4}$};
		\node [style=rn] (159) at (1.75, 2.25) {$\sigma_1\pi$};
		\node [style=rn] (160) at (2.75, 1.25) {$\sigma_2\pi$};
		\node [style=none] (161) at (-3, -2.5) {$=$};
	\end{pgfonlayer}
	\begin{pgfonlayer}{edgelayer}
		\draw (43.center) to (44.center);
		\draw (46.center) to (45.center);
		\draw [bend right=45] (55) to (58);
		\draw (107.center) to (108.center);
		\draw (110.center) to (109.center);
		\draw (113) to (114);
		\draw (113) to (115);
		\draw (113) to (116);
		\draw (114) to (111);
		\draw (115) to (112);
		\draw (119.center) to (120.center);
		\draw (122.center) to (121.center);
		\draw (125) to (128);
		\draw (123) to (125);
		\draw (124) to (125);
		\draw (132.center) to (133.center);
		\draw (135.center) to (134.center);
		\draw (138) to (141);
		\draw (136) to (138);
		\draw (137) to (138);
		\draw (55) to (38);
		\draw [bend left=45, looseness=0.75] (55) to (37);
		\draw (37) to (145);
		\draw (147) to (47);
		\draw (38) to (146);
		\draw (148) to (48);
		\draw (151.center) to (152.center);
		\draw (154.center) to (153.center);
		\draw [bend right=45] (157) to (158);
		\draw (157) to (150);
		\draw [bend left=45, looseness=0.75] (157) to (149);
		\draw (149) to (159);
		\draw (150) to (160);
		\draw (159) to (155);
		\draw (160) to (156);
	\end{pgfonlayer}
\end{tikzpicture}
} \end{center}

Above we have used both the $(f)$ and $(\pi)$ rules. We see now that we have ended up with a phase-gadget and we can re-write this in terms of a simple quantum circuit as pointed out in Appendix~\ref{app: ZX-calculus basics}.

\begin{center}
\scalebox{0.9}{
\begin{tikzpicture}
	\begin{pgfonlayer}{nodelayer}
		\node [style=none] (2) at (-5, -5.25) {};
		\node [style=none] (3) at (3.5, -5.25) {};
		\node [style=none] (4) at (-5, -4.25) {};
		\node [style=none] (5) at (3.5, -4.25) {};
		\node [style=gn] (6) at (-2.5, -4.25) {};
		\node [style=gn] (7) at (-0.75, -5.25) {};
		\node [style=rn] (8) at (-1.75, -6.25) {};
		\node [style=gn] (9) at (1.25, -4.25) {$(-1)^{\sigma_1}\frac{\pi}{4}$};
		\node [style=gn] (10) at (1.25, -5.25) {$(-1)^{\sigma_2}\frac{\pi}{4}$};
		\node [style=gn] (11) at (-2.25, -7.25) {$-\frac{\pi}{4}$};
		\node [style=rn] (12) at (-1, -4.25) {$(\sigma_1+\sigma_2)\pi$};
		\node [style=rn] (13) at (-4, -4.25) {$(\sigma_1+\sigma_2)\pi$};
		\node [style=none] (14) at (4, -4.75) {$=$};
		\node [style=none] (17) at (4.5, -5.25) {};
		\node [style=none] (18) at (14.75, -5.25) {};
		\node [style=none] (19) at (4.5, -4.25) {};
		\node [style=none] (20) at (14.75, -4.25) {};
		\node [style=gn] (23) at (12.5, -4.25) {$(-1)^{\sigma_1}\frac{\pi}{4}$};
		\node [style=gn] (24) at (12.5, -5.25) {$(-1)^{\sigma_2}\frac{\pi}{4}$};
		\node [style=rn] (25) at (10, -4.25) {$(\sigma_1+\sigma_2)\pi$};
		\node [style=rn] (26) at (5.5, -4.25) {$(\sigma_1+\sigma_2)\pi$};
		\node [style=rn] (27) at (7, -4.25) {};
		\node [style=rn] (28) at (8.5, -4.25) {};
		\node [style=gn] (29) at (7, -5.25) {};
		\node [style=gn] (30) at (8.5, -5.25) {};
		\node [style=gn] (31) at (7.75, -4.25) {$-\frac{\pi}{4}$};
	\end{pgfonlayer}
	\begin{pgfonlayer}{edgelayer}
		\draw (2.center) to (3.center);
		\draw (5.center) to (4.center);
		\draw (8) to (11);
		\draw (6) to (8);
		\draw (7) to (8);
		\draw (17.center) to (18.center);
		\draw (20.center) to (19.center);
		\draw (28) to (30);
		\draw (29) to (27);
	\end{pgfonlayer}
\end{tikzpicture}

}
\end{center}
We are using the ZX-calculus, but focusing on the specific circuit that is associated with the gadget. Thinking in this circuit scheme, we would like to have terms dependent on outcomes only on the right, such that they can be corrected using Clifford operations. Therefore, we will use the $(\pi)$ rule to propagate all $\sigma_i$-dependent gates to the right.\newline 
\begin{center}
\scalebox{0.9}{
\begin{tikzpicture}
	\begin{pgfonlayer}{nodelayer}
		\node [style=none] (2) at (-6, -0.5) {};
		\node [style=none] (3) at (3.75, -0.5) {};
		\node [style=none] (4) at (-6, 0.5) {};
		\node [style=none] (5) at (3.75, 0.5) {};
		\node [style=gn] (6) at (1.5, 0.5) {$(-1)^{\sigma_1}\frac{\pi}{4}$};
		\node [style=gn] (7) at (1.5, -0.5) {$(-1)^{\sigma_2}\frac{\pi}{4}$};
		\node [style=rn] (8) at (-0.5, 0.5) {$(\sigma_1+\sigma_2)\pi$};
		\node [style=rn] (9) at (-5, 0.5) {$(\sigma_1+\sigma_2)\pi$};
		\node [style=rn] (10) at (-3.5, 0.5) {};
		\node [style=rn] (11) at (-2, 0.5) {};
		\node [style=gn] (12) at (-3.5, -0.5) {};
		\node [style=gn] (13) at (-2, -0.5) {};
		\node [style=gn] (14) at (-2.75, 0.5) {$-\frac{\pi}{4}$};
		\node [style=none] (15) at (4, 0) {$=$};
		\node [style=none] (18) at (4.25, -0.5) {};
		\node [style=none] (19) at (13.75, -0.5) {};
		\node [style=none] (20) at (4.25, 0.5) {};
		\node [style=none] (21) at (13.75, 0.5) {};
		\node [style=gn] (22) at (12.25, 0.5) {$(-1)^{\sigma_1}\frac{\pi}{4}$};
		\node [style=gn] (23) at (12.25, -0.5) {$(-1)^{\sigma_2}\frac{\pi}{4}$};
		\node [style=rn] (24) at (10.25, 0.5) {$(\sigma_1+\sigma_2)\pi$};
		\node [style=rn] (25) at (6.5, 0.5) {$(\sigma_1+\sigma_2)\pi$};
		\node [style=rn] (26) at (5, 0.5) {};
		\node [style=rn] (27) at (8.75, 0.5) {};
		\node [style=gn] (28) at (5, -0.5) {};
		\node [style=gn] (29) at (8.75, -0.5) {};
		\node [style=gn] (30) at (8, 0.5) {$-\frac{\pi}{4}$};
		\node [style=none] (31) at (-6, -2.25) {$=$};
		\node [style=none] (34) at (-5.75, -2.75) {};
		\node [style=none] (35) at (5.75, -2.75) {};
		\node [style=none] (36) at (-5.75, -1.75) {};
		\node [style=none] (37) at (5.75, -1.75) {};
		\node [style=gn] (38) at (4.5, -1.75) {$(-1)^{\sigma_1}\frac{\pi}{4}$};
		\node [style=gn] (39) at (4.5, -2.75) {$(-1)^{\sigma_2}\frac{\pi}{4}$};
		\node [style=rn] (40) at (2.25, -1.75) {$(\sigma_1+\sigma_2)\pi$};
		\node [style=rn] (41) at (-0.75, -1.75) {$(\sigma_1+\sigma_2)\pi$};
		\node [style=rn] (42) at (-5.25, -1.75) {};
		\node [style=rn] (43) at (0.75, -1.75) {};
		\node [style=gn] (44) at (-5.25, -2.75) {};
		\node [style=gn] (45) at (0.75, -2.75) {};
		\node [style=gn] (46) at (-3.5, -1.75) {$(-1)^{\sigma_1+\sigma_2}(-\frac{\pi}{4})$};
		\node [style=none] (47) at (6.25, -2.25) {$=$};
		\node [style=none] (50) at (6.5, -2.75) {};
		\node [style=none] (51) at (13.75, -2.75) {};
		\node [style=none] (52) at (6.5, -1.75) {};
		\node [style=none] (53) at (13.75, -1.75) {};
		\node [style=gn] (54) at (12.5, -1.75) {$(-1)^{\sigma_1}\frac{\pi}{4}$};
		\node [style=gn] (55) at (12.5, -2.75) {$(-1)^{\sigma_2}\frac{\pi}{4}$};
		\node [style=rn] (58) at (7.25, -1.75) {};
		\node [style=rn] (59) at (11.25, -1.75) {};
		\node [style=gn] (60) at (7.25, -2.75) {};
		\node [style=gn] (61) at (11.25, -2.75) {};
		\node [style=gn] (62) at (9.25, -1.75) {$(-1)^{\sigma_1+\sigma_2}(-\frac{\pi}{4})$};
	\end{pgfonlayer}
	\begin{pgfonlayer}{edgelayer}
		\draw (2.center) to (3.center);
		\draw (5.center) to (4.center);
		\draw (11) to (13);
		\draw (12) to (10);
		\draw (18.center) to (19.center);
		\draw (21.center) to (20.center);
		\draw (27) to (29);
		\draw (28) to (26);
		\draw (34.center) to (35.center);
		\draw (37.center) to (36.center);
		\draw (43) to (45);
		\draw (44) to (42);
		\draw (50.center) to (51.center);
		\draw (53.center) to (52.center);
		\draw (59) to (61);
		\draw (60) to (58);
	\end{pgfonlayer}
\end{tikzpicture}
}
\end{center}
There are two distinct situations that can be considered:  (i) $\sigma_1\oplus \sigma_2 = 0$ or (ii) $\sigma_1\oplus \sigma_2 = 1$, where $\oplus$ represents addition modulo 2. In the first case, our task is already completed, that is, all the components that need to be corrected are on the right of the diagram, as we can use the fact that $(-1)^{\sigma_i}\frac{\pi}{4}= \frac{\pi}{4} - \frac{\sigma_i\pi}{2}$ and isolate the Clifford corrections from the non-Clifford gate $T$. On the other hand, when $\sigma_1 \oplus \sigma_2 = 1$, we use some small final diagrammatic manipulations to show that we will need to correct for an additional $CZ_{12}(S\otimes S)$ operator. \newline 
\begin{center}\scalebox{0.9}{
\begin{tikzpicture}
	\begin{pgfonlayer}{nodelayer}
		\node [style=none] (3) at (-7.75, -0.5) {};
		\node [style=none] (4) at (-1.5, -0.5) {};
		\node [style=none] (5) at (-7.75, 0.5) {};
		\node [style=none] (6) at (-1.5, 0.5) {};
		\node [style=gn] (7) at (-2.75, 0.5) {$(-1)^{\sigma_1}\frac{\pi}{4}$};
		\node [style=gn] (8) at (-2.75, -0.5) {$(-1)^{\sigma_2}\frac{\pi}{4}$};
		\node [style=rn] (9) at (-7, 0.5) {};
		\node [style=rn] (10) at (-4, 0.5) {};
		\node [style=gn] (11) at (-7, -0.5) {};
		\node [style=gn] (12) at (-4, -0.5) {};
		\node [style=gn] (13) at (-5.5, 0.5) {$\frac{\pi}{4}$};
		\node [style=none] (14) at (-1, 0) {$=$};
		\node [style=none] (17) at (-0.5, -0.5) {};
		\node [style=none] (18) at (6.5, -0.5) {};
		\node [style=none] (19) at (-0.5, 0.5) {};
		\node [style=none] (20) at (6.5, 0.5) {};
		\node [style=gn] (21) at (4.75, 0.5) {$\frac{\pi}{4}$};
		\node [style=gn] (22) at (4.75, -0.5) {$\frac{\pi}{4}$};
		\node [style=rn] (23) at (0.25, 0.5) {};
		\node [style=rn] (24) at (3.5, 0.5) {};
		\node [style=gn] (25) at (0.25, -0.5) {};
		\node [style=gn] (26) at (3.5, -0.5) {};
		\node [style=gn] (27) at (1.25, 0.5) {$-\frac{\pi}{4}$};
		\node [style=gn] (28) at (2.5, 0.5) {$\frac{\pi}{2}$};
		\node [style=gn] (29) at (5.75, 0.5) {$-\sigma_1\frac{\pi}{2}$};
		\node [style=gn] (30) at (5.75, -0.5) {$-\sigma_2\frac{\pi}{2}$};
		\node [style=none] (31) at (-5.25, -2) {$=$};
		\node [style=none] (34) at (-4.75, -2.5) {};
		\node [style=none] (35) at (5, -2.5) {};
		\node [style=none] (36) at (-4.75, -1.5) {};
		\node [style=none] (37) at (5, -1.5) {};
		\node [style=gn] (38) at (2.25, -1.5) {$\frac{\pi}{4}$};
		\node [style=gn] (39) at (2.25, -2.5) {$\frac{\pi}{4}$};
		\node [style=rn] (40) at (-4, -1.5) {};
		\node [style=rn] (41) at (-1.75, -1.5) {};
		\node [style=gn] (42) at (-4, -2.5) {};
		\node [style=gn] (43) at (-1.75, -2.5) {};
		\node [style=gn] (44) at (-3, -1.5) {$-\frac{\pi}{4}$};
		\node [style=gn] (45) at (0.5, -1.5) {$\frac{\pi}{2}$};
		\node [style=gn] (46) at (3.25, -1.5) {$-\sigma_1\frac{\pi}{2}$};
		\node [style=gn] (47) at (3.25, -2.5) {$-\sigma_2\frac{\pi}{2}$};
		\node [style=gn] (48) at (0.5, -2.5) {$\frac{\pi}{2}$};
		\node [style=gn] (49) at (-0.75, -1.5) {};
		\node [style=gn] (50) at (-0.75, -2.5) {};
		\node [style=none] (52) at (-5.25, -4) {$=$};
		\node [style=none] (55) at (-4.75, -4.5) {};
		\node [style=none] (56) at (5.5, -4.5) {};
		\node [style=none] (57) at (-4.75, -3.5) {};
		\node [style=none] (58) at (5.5, -3.5) {};
		\node [style=gn] (59) at (-0.75, -3.5) {$\frac{\pi}{4}$};
		\node [style=gn] (60) at (-0.75, -4.5) {$\frac{\pi}{4}$};
		\node [style=rn] (61) at (-4, -3.5) {};
		\node [style=rn] (62) at (-2, -3.5) {};
		\node [style=gn] (63) at (-4, -4.5) {};
		\node [style=gn] (64) at (-2, -4.5) {};
		\node [style=gn] (65) at (-3, -3.5) {$-\frac{\pi}{4}$};
		\node [style=gn] (66) at (2, -3.5) {$\frac{\pi}{2}$};
		\node [style=gn] (67) at (3.75, -3.5) {$-\sigma_1\frac{\pi}{2}$};
		\node [style=gn] (68) at (3.75, -4.5) {$-\sigma_2\frac{\pi}{2}$};
		\node [style=gn] (69) at (2, -4.5) {$\frac{\pi}{2}$};
		\node [style=gn] (70) at (0.75, -3.5) {};
		\node [style=gn] (71) at (0.75, -4.5) {};
		\node [style=hadamard] (72) at (-0.75, -2) {};
		\node [style=hadamard] (73) at (0.75, -4) {};
	\end{pgfonlayer}
	\begin{pgfonlayer}{edgelayer}
		\draw (3.center) to (4.center);
		\draw (6.center) to (5.center);
		\draw (10) to (12);
		\draw (11) to (9);
		\draw (17.center) to (18.center);
		\draw (20.center) to (19.center);
		\draw (24) to (26);
		\draw (25) to (23);
		\draw (34.center) to (35.center);
		\draw (37.center) to (36.center);
		\draw (41) to (43);
		\draw (42) to (40);
		\draw (49) to (50);
		\draw (55.center) to (56.center);
		\draw (58.center) to (57.center);
		\draw (62) to (64);
		\draw (63) to (61);
		\draw (70) to (71);
	\end{pgfonlayer}
\end{tikzpicture}
}\end{center}

Finally, the general form of the unitary injected by the constructed gadget is given, in ZX-calculus, by the diagram below.\newline 
\begin{center}\begin{tikzpicture}
	\begin{pgfonlayer}{nodelayer}
		\node [style=none] (3) at (-2.5, -1) {};
		\node [style=none] (5) at (-2.5, 0) {};
		\node [style=gn] (7) at (2.75, 0) {$\frac{\pi}{4}$};
		\node [style=gn] (8) at (2.75, -1) {$\frac{\pi}{4}$};
		\node [style=rn] (9) at (-1.25, 0) {};
		\node [style=rn] (10) at (0.75, 0) {};
		\node [style=gn] (11) at (-1.25, -1) {};
		\node [style=gn] (12) at (0.75, -1) {};
		\node [style=gn] (13) at (-0.25, 0) {$-\frac{\pi}{4}$};
		\node [style=none] (14) at (4.25, 0) {};
		\node [style=none] (15) at (4.25, -1) {};
		\node [style=none] (37) at (6.25, 0) {};
		\node [style=none] (38) at (6.25, -1) {};
		\node [style=none] (42) at (4.25, 0.5) {};
		\node [style=none] (43) at (6.25, 0.5) {};
		\node [style=none] (44) at (4.25, -1.5) {};
		\node [style=none] (45) at (6.25, -1.5) {};
		\node [style=none] (46) at (7.25, 0) {};
		\node [style=none] (47) at (7.25, -1) {};
		\node [style=none] (52) at (5.25, -0.5) {Corrections};
	\end{pgfonlayer}
	\begin{pgfonlayer}{edgelayer}
		\draw (10) to (12);
		\draw (11) to (9);
		\draw (3.center) to (15.center);
		\draw (5.center) to (14.center);
		\draw (44.center) to (42.center);
		\draw (42.center) to (43.center);
		\draw (43.center) to (45.center);
		\draw (45.center) to (44.center);
		\draw (37.center) to (46.center);
		\draw (38.center) to (47.center);
	\end{pgfonlayer}
\end{tikzpicture}\end{center} 
Three comments are in order. First, note that if we allow for the random application of either $T$ or $T^\dagger$ inside the \textsc{cnot} cascade structure, not caring which one is applied, there is no need for non-local Clifford corrections (i.e., the $CZ$ gates). Secondly, we point out that the white box labeled ``Corrections'' above corresponds to the transpose conjugate of the Clifford corrections presented in Eq.~\eqref{eq: Clifford corrections} of the main text. Finally, we see that we can make the gadget inject the state $\vert CS\rangle$ as we wanted.

\section{Proof of Lemma~\ref{lemma: new cat ent} }\label{app: entanglement lemma}

We want to show that $\mathcal{E}(\vert \text{cat}_m^* \rangle) = \frac{1}{2}$ for all $m\geq 2$. Recalling the definition of these cat states, we have that
\begin{align*}
    \vert \text{cat}^*_m\rangle \langle \text{cat}^*_m \vert &= \frac{1}{2}(\vert \text{cat}_m \rangle +e^{-i\pi/4}\vert \text{cat}_m^\perp\rangle)(\langle \text{cat}_m \vert + e^{i\pi/4}\langle \text{cat}_m^\perp \vert)\\ 
    &=\frac{1}{2}(\vert \text{cat}_m \rangle \langle \text{cat}_m \vert + e^{+i\pi/4}\vert \text{cat}_m \rangle \langle \text{cat}_m^\perp\vert +e^{-i\pi/4}\vert \text{cat}_m^\perp\rangle\langle \text{cat}_m\vert+\vert \text{cat}_m^\perp \rangle \langle \text{cat}_m^\perp \vert)
\end{align*}

If we call $\rho_k := \text{Tr}_{\setminus \mathcal{H}_k}(\vert \text{cat}^*_m\rangle \langle \text{cat}^*_m \vert)$, $\xi_k^{(1)} := \text{Tr}_{\setminus \mathcal{H}_k}(\vert \text{cat}_m\rangle \langle \text{cat}_m^\perp \vert)$ and $\xi_k^{(2)} := \text{Tr}_{\setminus \mathcal{H}_k} (\vert \text{cat}_m^\perp\rangle \langle \text{cat}_m \vert)$ we have that,
\begin{align*}
    \rho_k &= \frac{1}{2}(\text{Tr}_{\setminus \mathcal{H}_k}\vert \text{cat}_m \rangle \langle \text{cat}_m \vert + e^{+i\pi/4}\text{Tr}_{\setminus \mathcal{H}_k}\vert \text{cat}_m \rangle \langle \text{cat}_m^\perp\vert +e^{-i\pi/4}\text{Tr}_{\setminus \mathcal{H}_k}\vert \text{cat}_m^\perp\rangle\langle \text{cat}_m\vert+\text{Tr}_{\setminus \mathcal{H}_k}\vert \text{cat}_m^\perp \rangle \langle \text{cat}_m^\perp \vert)\\
    &=\frac{1}{2}\left(2\frac{\mathbb{1}_{2\times 2}}{2}+e^{+i\pi/4}\xi_k^{(1)}+e^{-i\pi/4}\xi_k^{(2)}\right) = \frac{\mathbb{1}_{2\times 2}}{2} + \frac{e^{+i\pi/4}}{2}\xi_k^{(1)} + \frac{e^{-i\pi/4}}{2}\xi_k^{(2)}
\end{align*}
where we have used the fact that $\text{Tr}_{\setminus \mathcal{H}_k}(\vert \text{cat}_m\rangle \langle \text{cat}_m \vert)= \text{Tr}_{\setminus \mathcal{H}_k}(\vert \text{cat}_m^\perp\rangle \langle \text{cat}_m^\perp \vert) = \mathbb{1}_{2\times 2}/2$. 

Since $\vert \text{cat}_m\rangle = \frac{1}{\sqrt{2}}(\vert T\rangle^{\otimes m} + \vert T^\perp \rangle^{\otimes m})$ and $\vert \text{cat}_m^\perp \rangle = \frac{1}{\sqrt{2}}(\vert T \rangle^{\otimes m} - \vert T^\perp \rangle^{\otimes m})$ we have that
\begin{equation*}
    \vert \text{cat}_m\rangle \langle \text{cat}_m^\perp \vert = \frac{1}{2}\left(\vert T\rangle^{\otimes m}\langle T\vert^{\otimes m}-\vert T\rangle^{\otimes m}\langle T^\perp \vert^{\otimes m}+\vert T^\perp \rangle^{\otimes m}\langle T\vert^{\otimes m}-\vert T^\perp\rangle^{\otimes m}\langle T^\perp\vert^{\otimes m}\right),
\end{equation*}
\begin{equation*}
    \vert \text{cat}_m^\perp \rangle \langle \text{cat}_m \vert = \frac{1}{2}\left(\vert T\rangle^{\otimes m}\langle T\vert^{\otimes m}+\vert T\rangle^{\otimes m}\langle T^\perp \vert^{\otimes m}-\vert T^\perp\rangle^{\otimes m}\langle T \vert^{\otimes m}-\vert T^\perp \rangle ^{\otimes m}\langle T^\perp \vert^{\otimes m}\right)
\end{equation*}
which implies that $\xi_k^{(1)}=\xi_k^{(2)}= \frac{1}{2}(\vert T \rangle \langle T \vert - \vert T^\perp \rangle \langle T^\perp \vert)$.

This means that we can write:
\begin{equation*}
    \rho_k = \frac{\mathbb{1}_{2\times 2}}{2} + \frac{1}{\sqrt{2}}\xi_k^{(1)} = \frac{1}{2} \left[ \left( 1 + \frac{1}{\sqrt{2}} \right)\vert T \rangle \langle T \vert + \left( 1 - \frac{1}{\sqrt{2}}\right)\vert T^\perp \rangle \langle T^\perp \vert \right]\,.
\end{equation*}
Taking the square of $\rho_k$ then leads to:
\begin{equation*}
    \rho_k^2 = \frac{1}{4} \left[ \left( 1 + \frac{1}{\sqrt{2}} \right)^2 \vert T \rangle \langle T \vert + \left( 1 - \frac{1}{\sqrt{2}}\right)^2 \vert T^\perp \rangle \langle T^\perp \vert \right]\,.
\end{equation*}

It is then straightforward to show that:
\begin{equation*}
    \text{Tr}(\rho_k^2) = \frac{3}{4},\forall k \implies \mathcal{E}(\vert \text{cat}_m^* \rangle )= 2 \left(1-\frac{1}{m}\sum_{k=0}^{m-1}\frac{3}{4}\right) = 2\left(1-\frac{3}{4}\right)=\frac{1}{2}
\end{equation*}
for all $m$-partite qubit states $\vert \text{cat}_m^*\rangle$.
\end{document}